\documentclass[11pt,a4paper]{article} 

\usepackage{adjustbox}
\usepackage{amsfonts}
\usepackage{amssymb}
\usepackage{amsmath}
\usepackage{amsthm}
\usepackage{authblk}
\usepackage{array}
\usepackage{bbm}
\usepackage{bigdelim}
\usepackage{booktabs}
\usepackage{caption}
\usepackage{subcaption}
\usepackage{dsfont}
\usepackage{enumerate}
\usepackage{framed,multirow}
\usepackage{graphics}
\usepackage{graphicx}
\usepackage{hyperref}
\usepackage{inputenc}
\usepackage{latexsym}
\usepackage{makecell}
\usepackage{mathrsfs}  
\usepackage[table]{xcolor}
\usepackage{tikz}
\usetikzlibrary{spy}
\usepackage[normalem]{ulem}
\usepackage{multirow}

\definecolor{ForestGreenOriginal}{rgb}{0.19, 0.73, 0.56}

\usepackage{arydshln} 
\newcommand{\centered}[1]{\begin{tabular}{l} #1 \end{tabular}} 

\usetikzlibrary{calc,fit}

\theoremstyle{definition}

\title{GEO-Nav: a geometric dataset of voltage-gated sodium channels}

\author[1]{Andrea Raffo}
\author[2,3]{Ulderico Fugacci}
\author[2,3]{Silvia Biasotti}

\affil[1]{Department of Biosciences, University of Oslo, 0316, Oslo, Norway}
\affil[2]{Istituto di Matematica Applicata e Tecnologie Informatiche  ``E. Magenes", Consiglio Nazionale delle Ricerche, 16149, Genova, Italy}
\affil[3]{RAISE Ecosystem, Genova, Italy}

\date{}                     
\newcolumntype{a}{>{\columncolor{blue!12}}m}
\newcolumntype{z}{>{\columncolor{teal!25}}m}

\begin{document}
\maketitle

\begin{abstract}
Voltage-gated sodium (Nav) channels constitute a prime target for drug design and discovery, given their implication in various diseases such as epilepsy, migraine and ataxia to name a few. In this regard, performing morphological analysis is a crucial step in comprehensively understanding their biological function and mechanism, as well as in uncovering subtle details of their mechanism that may be elusive to experimental observations. Despite their tremendous therapeutic potential, drug design resources are deficient, particularly in terms of accurate and comprehensive geometric information. This paper presents a geometric dataset of molecular surfaces that are representative of  Nav channels in mammals. For each structure we provide three representations and a number of geometric measures, including length, volume and straightness of the recognized channels. To demonstrate the effective use of GEO-Nav, we have tested it on two methods belonging to two different categories of approaches: a sphere-based and a tessellation-based method. \\
\textbf{Keywords}: dataset, sodium channels, molecular surfaces, geometric structures, computational biology
\end{abstract}

\section{Introduction}
\label{introduction}
Voltage-gated sodium (Nav) channels are transmembrane proteins consisting of a pore-forming pseudotetrameric $\alpha$ subunit and, in most vertebrate cells, one or smaller (non-pore-forming) $\beta$ subunits. Nav channels play an essential role in electrically excitable cells such as neurons, myocytes and endocrine cells: they respond to changes in \emph{membrane potential} -- the difference in electric potential between the interior and the exterior  of a cell -- by opening and closing ion-selective pores, allowing the flow of positively charged sodium ions from the region of higher concentration to the area of lower concentration (usually, from the exterior to the interior of the cell) \cite{Yu:2003}. It is worth noting that there is a background of deliberate ambiguity when using the word \emph{channel}. Not only is the term used for mentioning the whole protein (i.e., the transmembrane protein), but it also refers to a cavity -- i.e., a connected component of the complement space of the protein inside its convex hull -- with two entrances on opposite sides of a molecule (see, for example, \cite{Simoes:2017:STAR}).

\begin{figure}[h!]
	\centering
	\includegraphics[width=0.5\columnwidth]{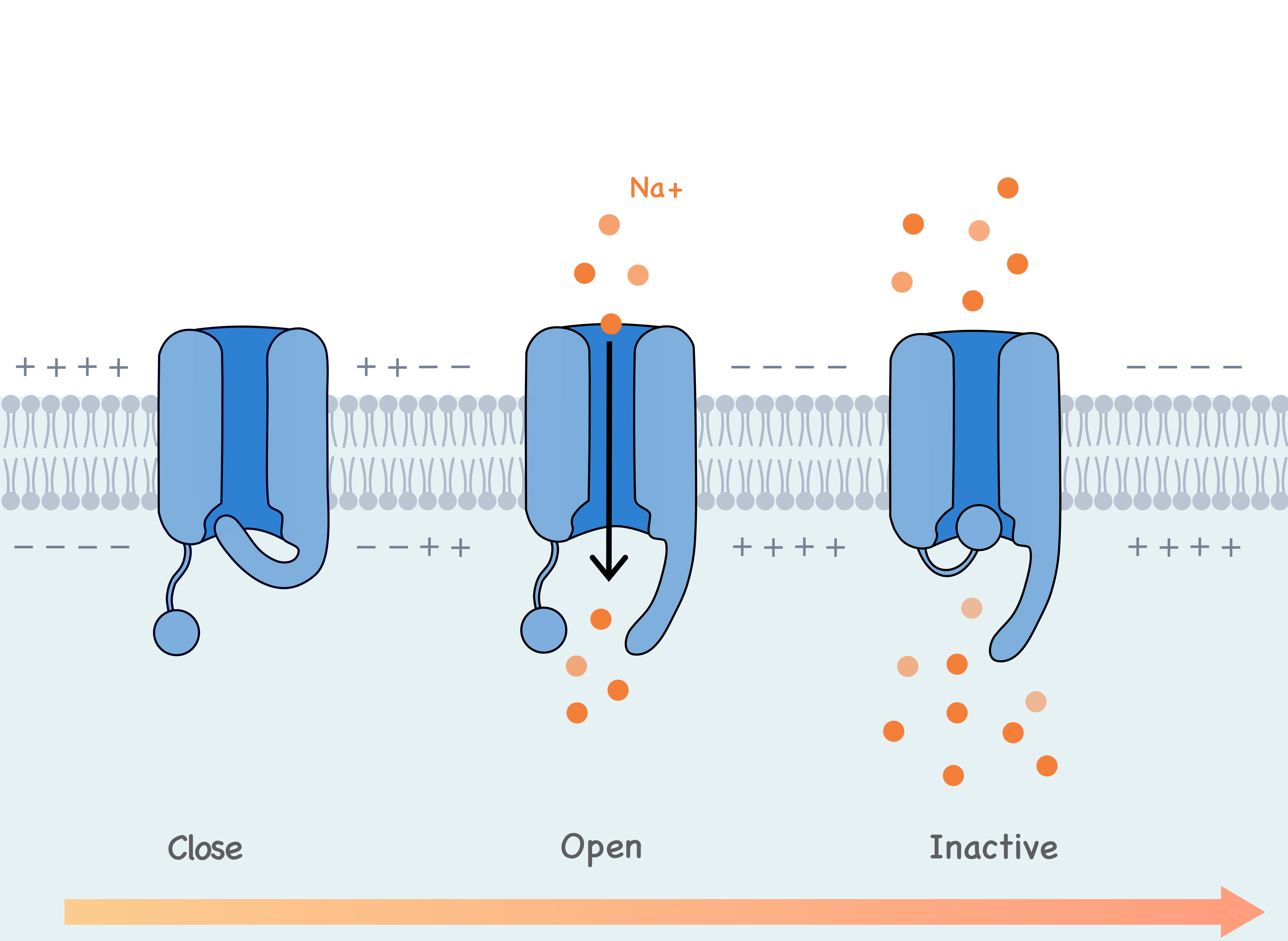}
	\caption{Three-state model of voltage-gated ion channels.}
	\label{fig:Nav_graph_illustration}
\end{figure}

\paragraph{Biological background and motivation}
There are three primary states in Nav channels, see \cite{deLeraRuiz:2015} and  Figure \ref{fig:Nav_graph_illustration}: 
\begin{itemize}
	\item \emph{Closed}. When the cell is at rest,  activation gates are closed, sodium ions are prevented from going through, and the membrane potential is maintained around some fixed negative value -- e.g., around -70 mV in most neurons.
	\item \emph{Open}. Because of an approaching action potential, the membrane potential rises (in jargon, it \emph{depolarizes}) -- e.g., to about -55 mV in  neurons; as a response,  activation gates open, allowing Na$^{+}$ ions to flow into the cell and the membrane potential to rise further.
	\item \emph{Inactivated}. As soon as the membrane potential has increased enough (to +30 mV in neurons), channels close their inactivation gates and the membrane becomes once more impermeable to Na$^{+}$ ions until the resting membrane potential is restored.
\end{itemize}

The prompt and precise opening and closing of voltage-gated sodium channels are essential for the generation and propagation of signals in electrically excitable tissues like nerves, hearts and muscles. Mutations in genes encoding Nav channels are among the major causes behind the development of neurological and psychiatric disorders (e.g., various types of idiopathic epilepsy and sensitivity to pain) \cite{Imbrici:2013}, cardiac diseases (e.g., Brugada syndrome, atrial fibrillation and dilated cardiomyopathy) \cite{Song:2012} and muscle disorders (e.g., potassium-aggravated myotonia and paramyotonia congenita) \cite{Simkin:2011}. Many drugs and toxins target voltage-gated sodium channels to alter their function, which can have therapeutic or harmful effects on the body.

Due to their relevance, the structure of sodium channels has been extensively studied using various techniques, including X-ray crystallography and cryo-electron microscopy. In parallel with in-vivo and in-vitro studies, structural bioinformatics has played an essential role in the study of sodium channels by providing insights into their atomic-level structures: it can be used to identify pores, approximate the channel centerlines\footnote{The line drawn between two sections of a tubular structure that maximize the distance from the boundary is here referred to as centerline.} and maximal inscribed ball radius values along them, quantify the channel size in terms of its volume and length, and distinguish between primary states. 

The survey \cite{Simoes:2017:STAR} classifies existing geometric algorithms into five main categories: (1) sphere-based, e.g., HOLE \cite{Smart:1996:HOLE}; (2) grid-based, such as Cavity search \cite{Ho:1990:Cavitysearch} and PoreWalker \cite{Pellegrini-Calace:2009:PoreWalker}; (3) surface-based, as for CHUNNEL \cite{Coleman:2009:CHUNNEL}; (4) tessellation-based, which includes MOLE \cite{Petrek:2007:MOLE}, MOLE 2.0 \cite{Sehnal:2013:MOLE2} and Chanalyzer \cite{Raffo:2022}; (5) methods combining the previous paradigms, e.g., CAVER and its offspring \cite{Petrek:2006:CAVER, Medek:2008:CAVER2, Chovancova:2012:CAVER3}.

\paragraph{Contribution} The main contributions of this paper include:
\begin{itemize}
	\item The first geometry-oriented dataset of Nav channels, created from a set of $21$ structures and additional $63$ synthetically-generated ones. Three additional synthetic models are made available together with the ground truth values they were constructed with -- to perform initial testing of algorithms and quantitatively estimate the outcome of a method with the geometry of a channel.  Each structure is made available both as a PDB file and as its Solvent Excluded Surface (SES) -- generated by the freely available software NanoShaper \cite{DeCherchi2013}.
	\item The output of the in-silico characterizations obtained using two computational methods: the sphere-based method HOLE \cite{Smart:1996:HOLE} and the tessellation-based tool Chanalyzer \cite{Raffo:2022}.
	\item A number of routines to study the centerlines and maximal inscribed ball radius values along them: centerline length and straightness, and channel volume. Two comparison measures are also provided.
	\item Quantitative and qualitative analyses of the output, together with a comparison that highlights the strengths and weaknesses of the two approaches.
\end{itemize}

\paragraph{Organization} The remainder of the paper is organized as follows. Section \ref{sec:related_datasets} briefly discusses existing databases and details how this dataset differs from those presented in the previous editions of the Symposium of 3D Object Retrieval. Section \ref{dataset} describes how the dataset was generated and what files are available. Two geometric algorithms, summarized in Section \ref{sec:methods}, are evaluated and compared on the GEO-Nav dataset on the basis of measures presented in Section \ref{sec:measures}; results and considerations derived from such an analysis are presented in  Section \ref{sec:comparison}. Concluding remarks end the paper.

\section{Related datasets and databases\label{sec:related_datasets}}
Datasets presented to the previous editions of the Symposium of 3D Object Retrieval \cite{shrec2017,shrec18,shrec2019,shrec2020,Raffo:2021,Gagliardi:2022} focused on the retrieval, classification and segmentation of small molecules with no restriction on their function. We here consider voltage-gated sodium channels, macromolecules counting approximately 2000 amino acid residues organized in four homologous domains, and a detailed description of the geometry of each channel -- including centerline, volume, length and straightness.

Although some comprehensive databases, such as PubChem \cite{Kim:2021:PubChem}, ChEMBL \cite{Mendez:2019:ChEMBL}, BindingDB  \cite{Gilson:2016:BindingDB} and VGSC-DB \cite{Wang:2022:VGSC-DB} collect several relevant data on Nav channels, they are either general-purpose (i.e., they do not consider just channels but the more general case of cavities) or their focus restricts to the sole phisico-chemical information, neglecting the geometry of the problem. Moreover, such databases do not provide any sort of measure to test and compare the capability of computational approaches to study channels. 

\section{The dataset}
\label{dataset}

\begin{figure*}[h!]
	\centering
	\includegraphics[width=0.7\textwidth]{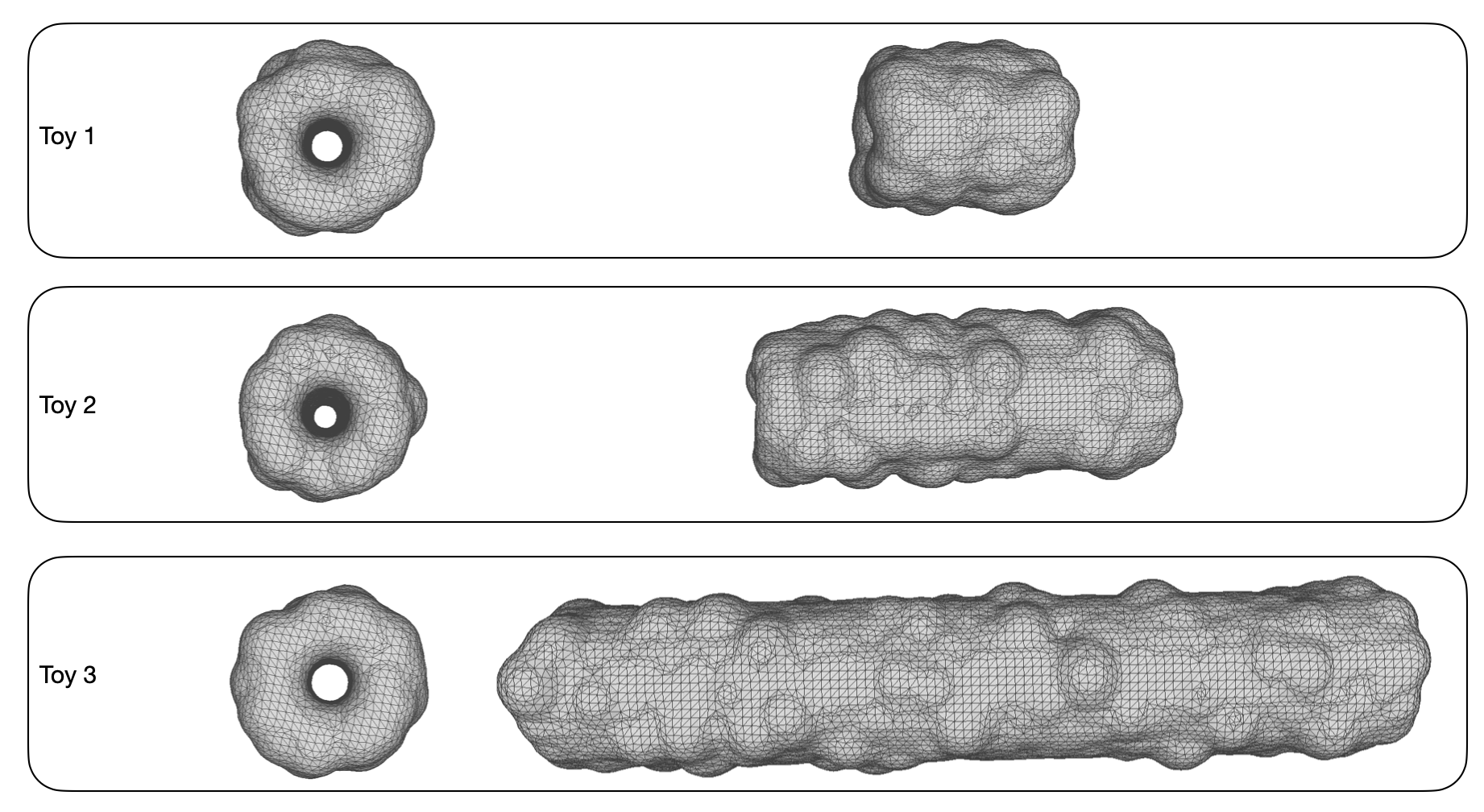}
	\caption{Molecular surfaces of three benchmarked models. The molecular surfaces were generated with NanoShaper and visualized in MeshLab \cite{Cignoni:2008}.}
	\label{fig:OFFs_toys}
\end{figure*}

Recognizing and characterizing voltage-gated sodium channels are key steps toward unravelling their intricate mechanisms and their involvement in various physiological and pathological conditions.
The construction of a reliable and comprehensive dataset of sodium channels necessitates the output of domain experts, which is here incorporated by following some ground properties of what such channels are supposed to look like. The dataset here presented consists of: 
\begin{itemize}
	\item $3$ \emph{benchmarked (BM)} models, synthetically created by placing carbons around straight segments so that the cavity resembles a cylinder of fixed radius (see Figure \ref{fig:OFFs_toys}). While unrealistic if compared to voltage-gated sodium channels, they can be used as a sanity check given that their centerline and maximal inscribed ball radius values are known a priori (by construction). These $3$ models are meant to differ from Nav channels for which some crystallographic structure (or molecular dynamics) may be known, but, at each instant, only the experimental estimate of the position of the atoms at a given instant is available.
	\item $21$ PDB entries deposited in the Protein Data Bank archive \cite{PDB2000}, each one containing a Nav channel in different conformations experimentally determined via X-ray diffraction and cryo-electron microscopy. The structures refer to eight Nav channels: Nav1.1-Nav1.8. Mutations differentially expressed throughout the human body are associated with migraine (Nav1.1), epilepsy (Nav1.1–Nav1.3, Nav1.6), cardiac (Nav1.5), pain (Nav1.7–Nav1.8), and muscle paralysis (Nav1.4) syndromes \cite{Clairfeuille:2017}. A complete list is provided in Table \ref{tab:PBDs}, together with details about the specific structure and its resolution (a measure of the quality of the structure, which quantifies how much detail can be observed in the experimental data). The existence of a cavity that qualifies as a channel is certified by the metadata contained in the PDB repository as well as in the  papers that introduced and analyzed these structures -- which are listed in Table \ref{tab:PBDs}.
	\item $63$ additional structures, three for each of PDB entries of the previous point. Each structure is obtained by applying a uniform random perturbation in $[-1,1]^3$ $\\A$ to the coordinates of those atoms that are sufficiently farther away from the centerline than a given threshold. Since we can't determine the centerline's exact location without making use of computational methods, we proceed manually by temporarily roto-translating the channel (here intended as the whole protein) as if the centerline is approximately aligned to the $z$-axis; we then consider a threshold distance of $10\AA$ from the $z$-axis. Although this procedure might sound way overly subjective and prone to serious human error, a threshold distance of $10\AA$ is sufficiently large to leave atoms determining the cavity untouched. It is indeed known that in Navs channels -- more generally, in transmembrane proteins -- channels have preferably straight centerlines \cite{Masood:2015}; additionally, it has been observed that Navs have maximal inscribed ball radius values roughly in 3.6-6 $\AA$, with average bottleneck radii ranging from 1.62 to 2.20 $\AA$  \cite{Kaczmarski:2014, Ullrich:2021}.
\end{itemize}
This leads to a total of $84$ structures, which are made available at the link \url{https://github.com/rea1991/GEO-Nav} in three formats: PDB files, XYZR files, and OFF\footnote{\url{https://segeval.cs.princeton.edu/public/off\_format.html}} files.

PDB files are required by many of the tools from the bioinformatics community, e.g., from HOLE \cite{Smart:1996:HOLE}. Before use, PDB files need to be preprocessed using a custom, freely-distributed, Python script\footnote{\url{https://electrostaticszone.eu/downloads/scripts-and-utilities/15-pdb2xyzr.html}} to remove ligands and Heterogeneous atoms (HETATM). 

\begin{figure*}[h!]
	\centering
	\begin{tabular}{ccc}

		\begin{tikzpicture}[spy using outlines={circle,black,magnification=3.8,size=2.75cm, connect spies}]
			\node {\includegraphics[width=4.75cm,trim={16cm 7cm 15cm 4cm}, clip]{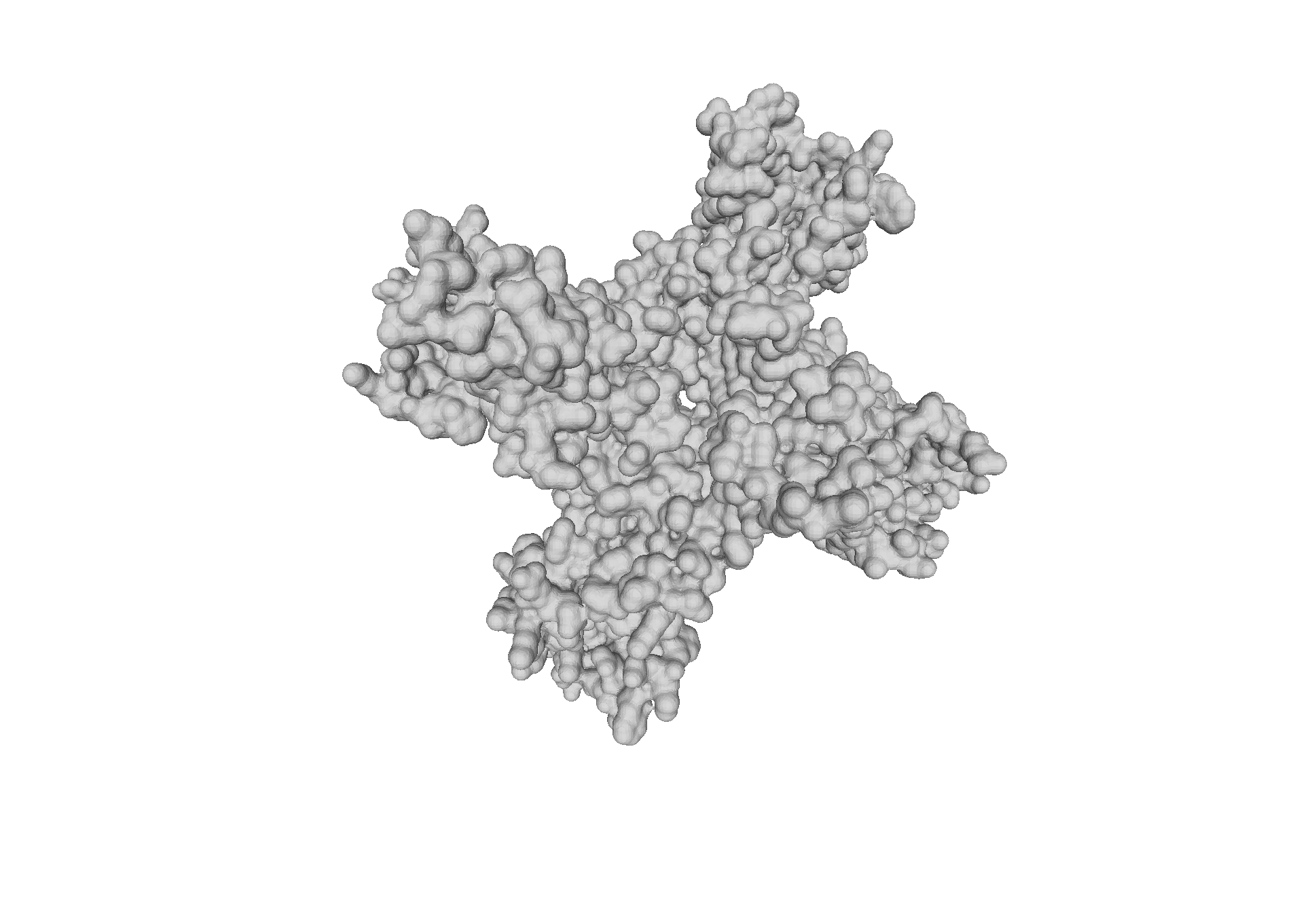}};
			\spy on (0.15,0.15) in node [left] at (2.,-3.5);
		\end{tikzpicture}
		&
		\begin{tikzpicture}[spy using outlines={circle,black,magnification=3.8,size=2.75cm, connect spies}]
			\node {\includegraphics[width=4.75cm,trim={16cm 5cm 12cm 2cm}, clip]{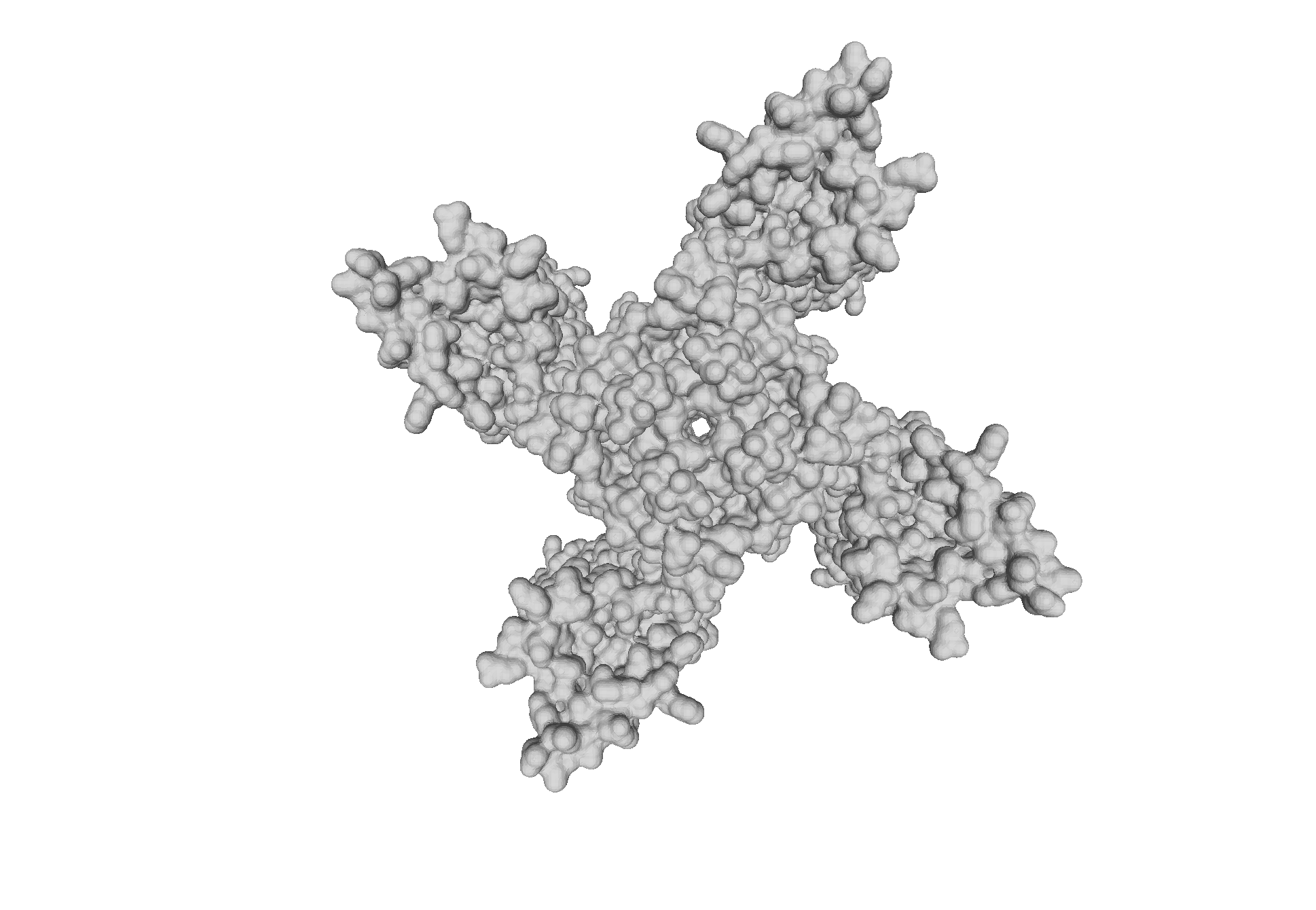}};
			\spy on (0.025,0.0) in node [left] at (2.,-3.5);
		\end{tikzpicture}
		&

		\begin{tikzpicture}[spy using outlines={circle,black,magnification=3.8,size=2.75cm, connect spies}]
			\node {\includegraphics[width=4.75cm,trim={16cm 5cm 16cm 4cm}, clip]{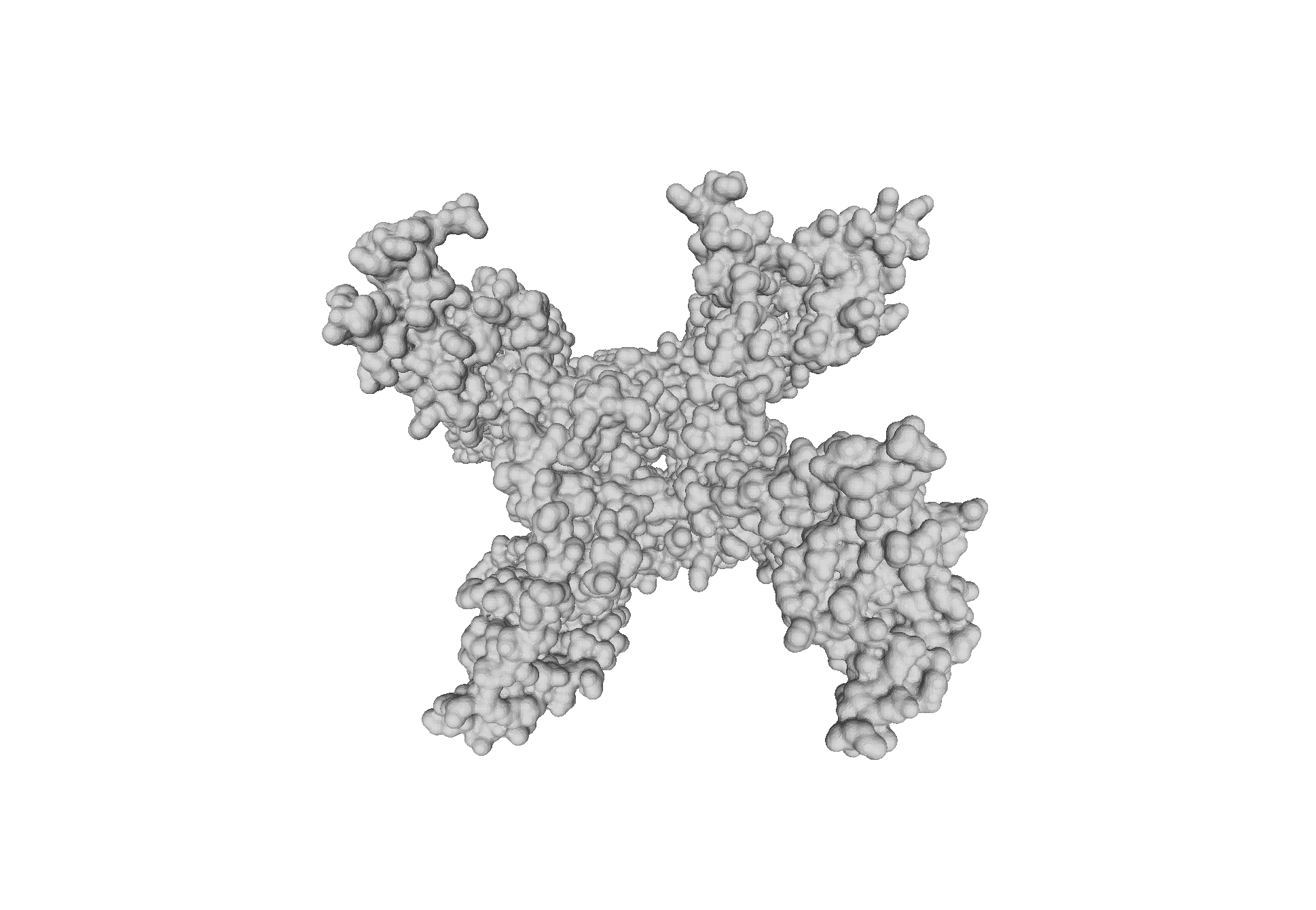}};
			\spy on (0.04,-0.15) in node [left] at (2.,-3.5);
		\end{tikzpicture}
		\\ (a) & (b) & (c) 
		\\
		
	\end{tabular}
	\caption{Examples of molecular surfaces generated with NanoShaper and visualized in MeshLab \cite{Cignoni:2008}: (a) 6LQA, (b) 7K48, and (c) 8FHD.\\}
	\label{fig:OFFs_graph_ab}
\end{figure*}

OFF files contain the molecular surfaces (MS) calculated and triangulated with the C++ sofware NanoShaper \cite{DeCherchi2013}, choosing the Connolly Solvent Excluded Surface model \cite{Richards1977, Connolly}; default parameters are considered, with the sole exception of the probe radius which is set to $0.8\AA$. To produce OFF files, NanoShaper requires XYZR files: an XYZR file lists one atom per row, while columns contain the coordinates of the atomic centers and the atomic radii. To convert PDB files to XYZR files we have used a Python script based on the open-source package \texttt{ProDy}\footnote{\url{http://prody.csb.pitt.edu}} for protein structural dynamics analysis and on the Amber99 force field \cite{Ponder:2003:force_fields}. Examples of protein surfaces produced by NanoShaper are provided in Figure \ref{fig:OFFs_graph_ab}.

\begin{table*}[h!]      
	\caption{List of the $21$ PDB entries selected from the Protein Data Bank repository. For each PDB code, we report: the protein name, general details on the structure, the resolution it has been acquired, and the reference paper where the channel was described.}
	\begin{center}
			\resizebox{\columnwidth}{!}{
			\begin{tabular}{|c|c|p{8cm}|c|c|}
				\hline
				\cellcolor{ForestGreenOriginal!35}PDB code 
				&
				\cellcolor{ForestGreenOriginal!35}Protein name
				&
				\cellcolor{ForestGreenOriginal!35}{\;\;\;\;\;\;\;\;\;\;\;\;\;\;\;\;\;\;\;\;\;\;\;\;\;\;\;\;\;\;\;\;\;\;\;}Details 
				&  
				\cellcolor{ForestGreenOriginal!35}Resolution (\r{A})
				&  
				\cellcolor{ForestGreenOriginal!35}Reference
				\\
				\hline
				5EK0 & Nav 1.7 & Chimera\footnote{Chimera (or chimeric, fusion) proteins are proteins obtained by fusing two or more genes that originally coded for separate proteins} of bacterial ion transport protein and human sodium channel protein & 3.53 & \cite{Ahuja:2015} \\
				5XSY & Nav1.4 & Complex with beta1 subunit from Electrophorus electricus (electric eel)  & 4.00 & \cite{Yan:2017} \\
				6AGF & Nav1.4 & Human, in complex with beta1  & 3.20 & \cite{Pan:2018} \\
				6J8E & Nav1.2 & Human, into a ternary complex with beta2-KIIIA & 3.00 & \cite{Pan:2019} \\
				6J8G & Nav1.7 & Human, ternary complex with beta1-beta2 and bound to two combinations of pore blockers and gating modifier toxins, huwentoxin-IV and saxitoxin & 3.20 & \cite{Shen:2019} \\
				6J8H & Nav1.7 & Human, in complex with auxiliary beta1-beta2 subunits, huwentoxin-IV and saxitoxin & 3.20 & \cite{Shen:2019} \\    
				6J8J & Nav1.7 & Human, in complex with auxiliary beta1-beta2 subunits, ProTx-II and tetrodotoxin & 3.20 & \cite{Shen:2019} \\    
				6LQA & Nav1.5 & Human, with the antiarrhythmic agent quinidine & 3.40 & \cite{Li:2021:1} \\
				7DTC & Nav1.5 & Human, with E1784K mutation (associated with Brugada syndrome)  & 3.30 & \cite{Li:2021:2} \\
				7DTD & Nav1.1 & Human, in complex with beta4 & 3.30 & \cite{Pan:2021} \\
				7K18 & Nav1.5 & Cardiac channel with toxin bound  & 3.30 & \cite{Jiang:2021}\\
				7K48 & Nav1.7 & NavAb/Nav1.7-VS2A chimera trapped in the resting state by tarantula toxin m3-Huwentoxin-IV & 3.60 & \cite{Wisedchaisri:2021} \\
				7W7F & Nav1.3 & Human NaV1.3/beta1/beta2-ICA121431 & 3.35 & \cite{Li:2022}\\
				7W9K & Nav1.7 & Complex with beta1-beta2 subunits & 2.20 & \cite{Huang:2022:1} \\
				7W77 & Nav1.3 & Complex with beta1/beta2 subunits and Bulleyaconitine A & 3.30 & \cite{Li:2022}\\
				7WE4 & Nav1.8 & Human, with the blocker A-803467, class I & 2.20 & \cite{Huang:2022:2} \\
				7WEL & Nav1.8 & Human, with the blocker A-803467, class II & 3.20 & \cite{Huang:2022:2} \\
				7WFR & Nav1.8 & Human, with the blocker A-803467, class III & 3.00 & \cite{Huang:2022:2}\\
				7WFW & Nav1.8 & Human  & 3.10 & \cite{Huang:2022:2}\\
				7XMF & Nav1.7 & Human, in complex with beta1/beta2 subunits and Nav1.7-IN2 (channel inhibitor) & 3.07 & \cite{Zhang:2022}\\
				8FHD & Nav1.6 & Human, in complex with beta1 subunit & 3.10 & \cite{Fan:2023} \\
				\hline 
			\end{tabular}
		}
		\label{tab:PBDs}
	\end{center}
\end{table*}

\section{Evaluation measures}\label{sec:measures}

When applying a geometric method to a PDB or OFF file containing a channel, the expected output consists of a collection of points representing the centerline of the Nav channel, together with the maximal inscribed ball radius values at such points. Specifically, each point $\mathbf{p}$ of the centerline $C$ is endowed with its coordinates and a value $\rho_\mathbf{p}$ representing the radius of the largest sphere centred at the considered point $\mathbf{p}$ and completely contained in the Nav channel.

We develop a collection of elementary tools which, given a molecular structure, aim at comparing the protein channel recognized by different computational methods in terms of centerlines and maximum inscribed ball radius values. 

\subsection{Measures for single-channel analysis}
We are able to perform some elementary measures of the retrieved channels such as the length, the straightness, and the volume. 

Assuming that the points of a centerline $C$ are sorted from one entrance $\mathbf{p}_0$ of the channel to the other entrance $\mathbf{p}_N$, the length value is simply retrieved by summing the distances between consecutive points of $C$:
\begin{equation*}
	L(C):=\sum_{j=1}^N\|\mathbf{p}_j-\mathbf{p}_{j-1}\|_2.
\end{equation*}

The \emph{straightness} of a centerline $C$ is measured as the reciprocal of the \emph{tortuousness} of $C$, i.e.,
\begin{equation*}
	s(C):=\dfrac{1}{tortuousness(C)},
\end{equation*}
where the tortuousness is defined as the average distance of the points of $C$ to the straight line $\ell$ passing through the first and the last point of $C$; the tortuousness of the centerline of a channel is a positive real number close to zero when $C$ is rectilinear and assumes greater values as the path loses its rectilinear behaviour. As a consequence, high straightness values denote almost rectilinear centerlines, while low values notify a more curvilinear behaviour.

Lastly, the volume of a channel can be simply retrieved by computing the volume of the region of space obtained as the union, for $\mathbf{p}$ varying in $C$, of the balls $B(\mathbf{p},\rho_\mathbf{p})$ of radius $\rho_\mathbf{p}$ centred at $\mathbf{p}$. In formulas, this corresponds to
\begin{equation*}
	V(C,\rho):=\iiint_{\bigcup_{\mathbf{p}\in{}C}B(\mathbf{p},\rho_\mathbf{p})}1dxdydz.
\end{equation*}

\subsection{Comparison measures}
Interesting information arises from comparing the output of different computational methods. Such an analysis can involve channels of different molecular structures or -- as it will be more detailed discussed in Section \ref{sec:comparison} -- channels recognized on the same molecule by different software tools.

A preliminary comparison between centerlines can be achieved by considering some of the previously introduced measures such as the length, the straightness, and the volume.

To make the comparison between channel centerlines more effective, we are interested in identifying portions of the centerlines common to both channels. This identification is measured by the function $match$. Given two centerlines $C$ and $C'$, we consider a point $\mathbf{p} \in C$ matched with $C'$ if there exists a point $\mathbf{p}' \in C'$ such that the Euclidean distance between $\mathbf{p}$ and $\mathbf{p}'$ is below a certain threshold value. We define $match(C, C')$ as the percentage of points of $C$ matched with $C'$.

Another comparison measure between two centerlines allows for evaluating the difference between the radius values assumed by the centerlines on their portion identified as matched. Specifically, the matching between two centerlines $C$ and $C'$ defines a correspondence $\gamma: M \rightarrow M'$ between two portions $M$ and $M'$ of $C$ and $C'$, respectively. The distance $d_\rho(C,C')$ between the radius values of $C$ and $C'$ is defined, analogously to the $L^2$-norm, as
$$d_\rho(C,C'):=\sqrt{\sum_{\mathbf{p} \in M}| \, \rho_{\mathbf{p}} - \rho_{\gamma(\mathbf{p})} \, |^2},$$
where we recall that $\rho_\mathbf{p}$ denotes the largest sphere centered at $\mathbf{p}$ completely contained in the considered channel.

\section{Methods\label{sec:methods}}
The proposed GEO-Nav dataset has been used to evaluate and compare two methods for the geometric study of voltage-gated sodium channels: HOLE and Chanalyzer.

\subsection{HOLE}
The software \cite{Smart:1996:HOLE}, implemented in FORTRAN-77, is de facto one of the standards in the analysis of static transmembrane channel proteins. HOLE relies on user-assisted cavity localization (UACL), i.e., it requires an initial point $\mathbf{p}$, that lies anywhere within the channel, and a vector $\mathbf{v}$, that is approximately in the direction of the channel. 

The initial point is firstly adjusted by the Metropolis Monte Carlo simulated annealing procedure to find the sphere satisfying the following two requirements: (i) its center lies on the plane determined by $\mathbf{p}$ and $\mathbf{v}$; (ii) it is the largest  that does not overlap with the atoms bordering the channel. Here, an atom is regarded as a sphere of given center and with radius equal to the element's van der Waals radius  \cite{Weiner:1984:AMBER}.

Such a sphere -- regarded as a probe sphere -- is then rolled through the channel, while its radius is adjusted by the Metropolis Monte Carlo simulated annealing procedure. In practice, this is repeated by taking a small displacement in the direction of the vector $\mathbf{v}$ until the end of the channel has been reached: this happens when the accommodated sphere radius exceeds a user-defined value, whose default value is $5Å$. 

The process is restarted from $\mathbf{p}$ using the vector $-\mathbf{v}$ as direction. To properly sort the points of the centerline $C$, we develop and perform an algorithm based on tools of graph theory capable of interpreting $C$ as a path from one entry of the channel to the other.

\subsection{Chanalyzer}
Chanalyzer \cite{Raffo:2022} is a tool designed to detect and geometrically describe ion channels in molecular dynamics trajectories through methods from computational geometry. 
Chanalyzer proceeds in four steps, which are here summarized to give a general idea of the working principles: (1) extraction of the tetrahedral representation of the channel via the alpha shape theory; (2) projection of the tetrahedral representation of the channel onto the SES generated via NanoShaper; (3) skeletonization and extraction of source and target points; (4) centerline computation. The four steps are applied subsequently without the need for manual interaction.

\paragraph{Step 1: extraction of the tetrahedral representation of the channel} Starting from a molecule described as a collection of three-dimensional balls, Chanalyzer builds its weighted Delaunay triangulation; then, it filters the simplices to discard those tetrahedra that (i) belongs to the \emph{dual complex}, i.e., the set of simplices that are completely inside the molecule, or (ii) are ancestors of the complement of the convex hull with respect to a discrete-flow, which works according to the principle of a fluid flowing into a sink. The connected component having the largest volume corresponds to the volumetric approximation of the channel. 

\paragraph{Step 2: projection of the tetrahedral representation of the channel onto the molecular surface} After computing a triangular mesh approximation of the Solvent Excluded Surface via the software NanoShaper, a portion of the SES that roughly corresponds to the channel is extracted by discarding all those triangles having their barycenter outside the tetrahedral approximation.

\paragraph{Step 3: skeletonization and extraction of source and target points} The mean curvature flow skeleton of the sub mesh -- which corresponds to an approximation of the medial axis of the channel -- is computed via the Computational Geometry Algorithms Library (CGAL \cite{CGAL}). Chanalyzer then extracts an approximated centerline of the channel by pruning the just-computed skeleton; it proceeds  by maximizing a score that takes into account both the length of a candidate centerline and its rectilinear behaviour.   

\paragraph{Step 4: centerline computation} The two ends of the approximated centerline serve as source and target points of the Vascular Molecular ToolKit (VMTK \cite{Piccinelli:2009:VMTK}), a standard software package that can produce a more accurate version of the centerline. For a given molecular surface, the output of Chanalyzer includes the centerline as a list of three-dimensional points and the list of maximal inscribed ball radius values along them.



\section{Experimental analysis}\label{sec:comparison}

As a guiding example of the use of the dataset described in Section \ref{dataset} and evaluation measures introduced in Section \ref{sec:measures}, we run here the tools HOLE and Chanalyzer described in Section \ref{sec:methods} and compare their output. 

As previously mentioned, the process of user-assisted cavity localization adopted by HOLE requires the user to provide the initial point and direction of the channel, which are then adjusted by a Metropolis Monte Carlo simulated annealing procedure. In order to better compare the two software tools, we have run HOLE twice for each structure of the dataset providing as initial information the two initial points and directions retrieved as the entrances of the channel by Chanalyzer; this has, additionally, avoided any need for manual interaction.

\begin{figure*}[h!]
	\begin{center}
		\resizebox{0.36\textheight }{!}{	
			\begin{tabular}{c  c  c  c  c  c  c } 
				&  & BM 1 & & BM 2 & & BM 3\\
				\rotatebox[origin=c]{90}{ Centerline Chanalyzer}&  &
				\centered{\includegraphics[width=0.16\linewidth]{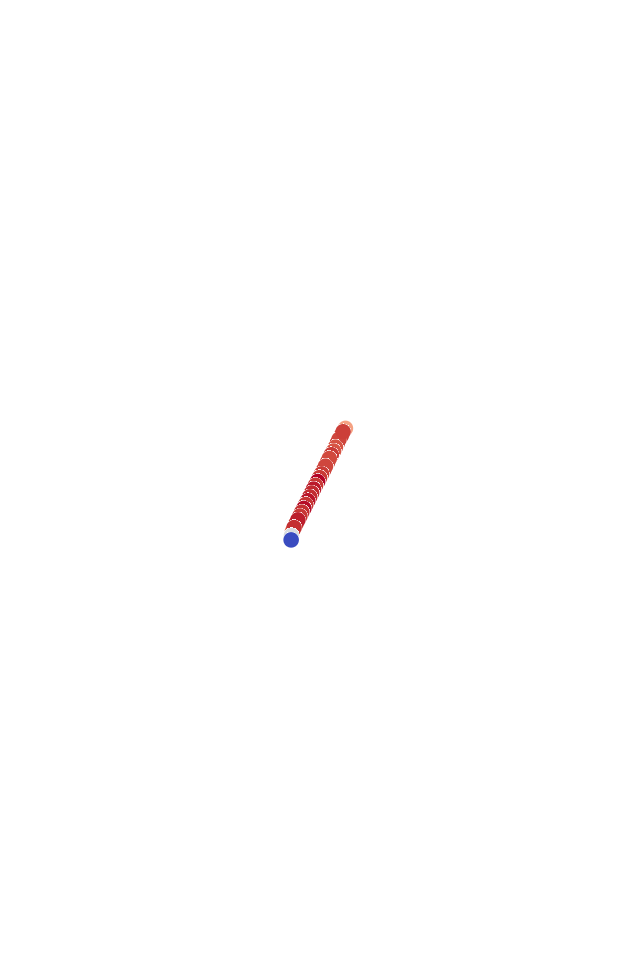} }& & \centered{\includegraphics[width=0.16\linewidth]{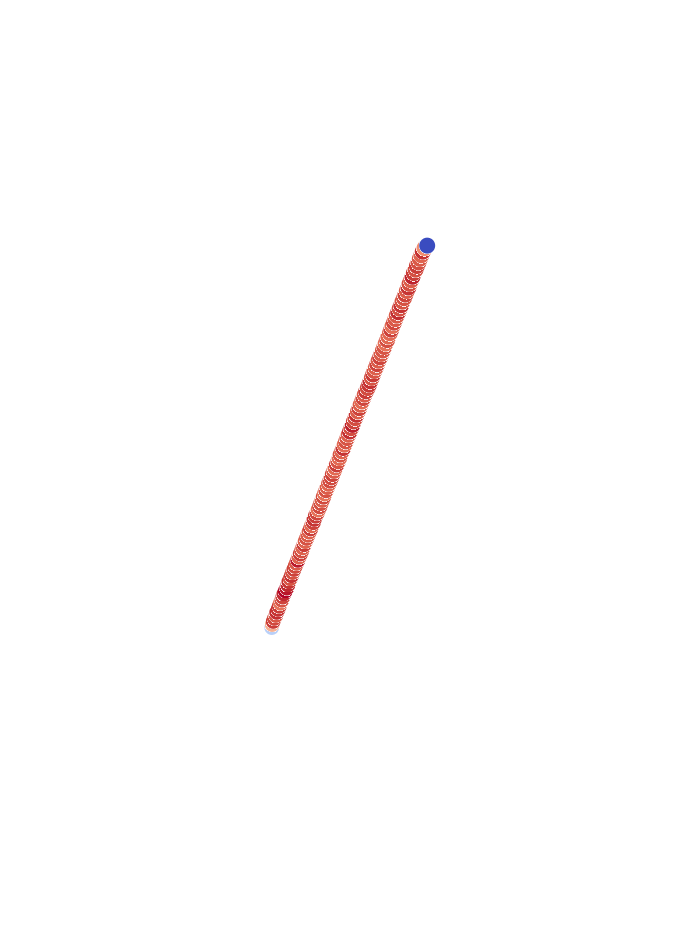}} & &
				\centered{\includegraphics[width=0.16\linewidth]{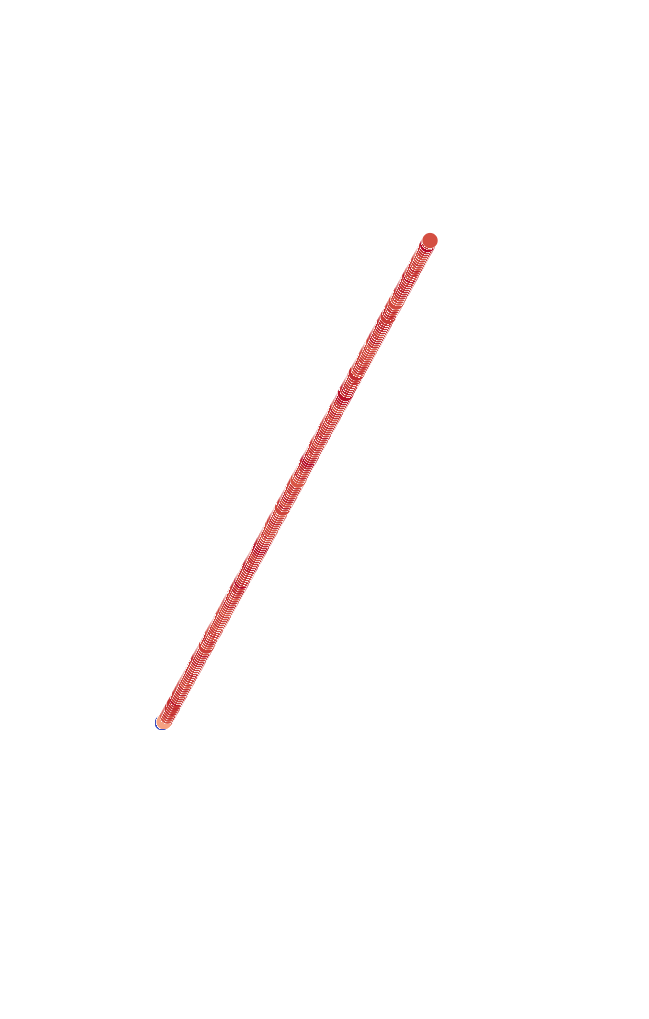}}\\
				\rotatebox[origin=c]{90}{Centerline 1 HOLE} &  &
				\centered{\includegraphics[width=0.16\linewidth]{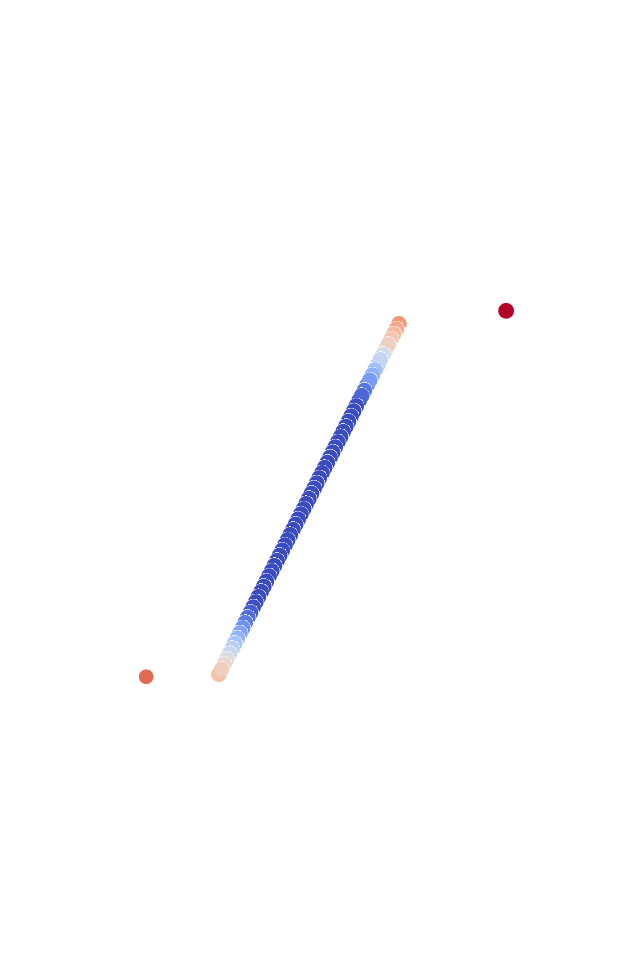}} & &\centered{\includegraphics[width=0.16\linewidth]{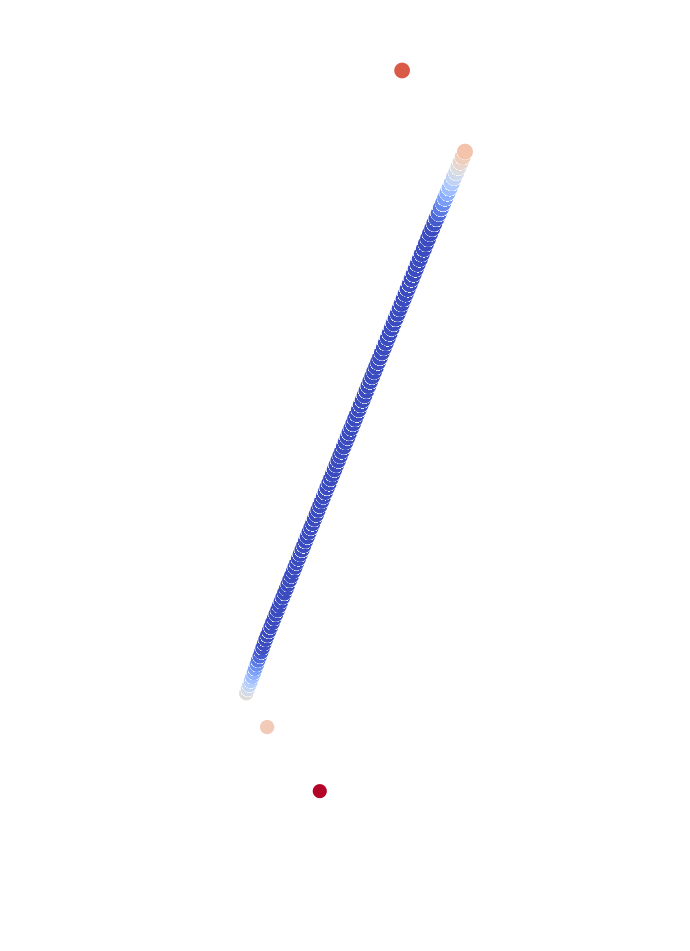}} & &
				\centered{\includegraphics[width=0.16\linewidth]{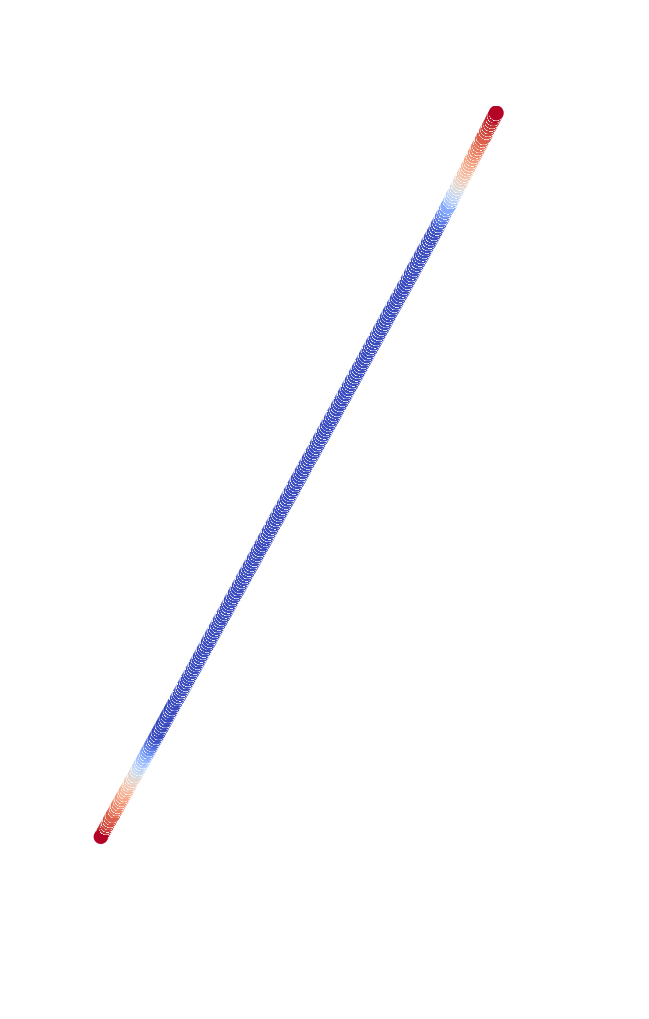}}\\
				\rotatebox[origin=c]{90}{Centerline 2 HOLE} &  &
				\centered{\includegraphics[width=0.16\linewidth]{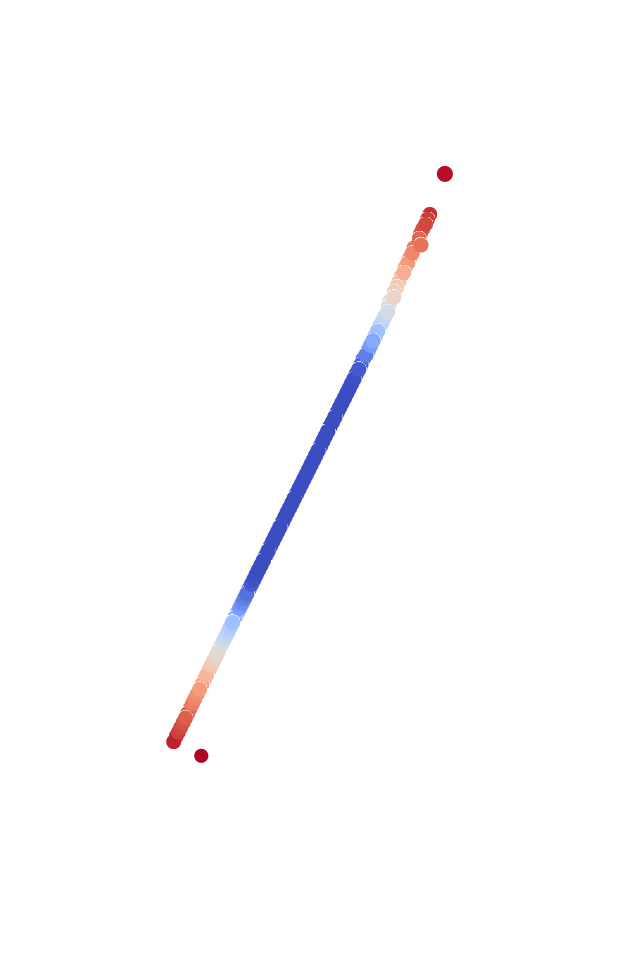}} & & \centered{\includegraphics[width=0.16\linewidth]{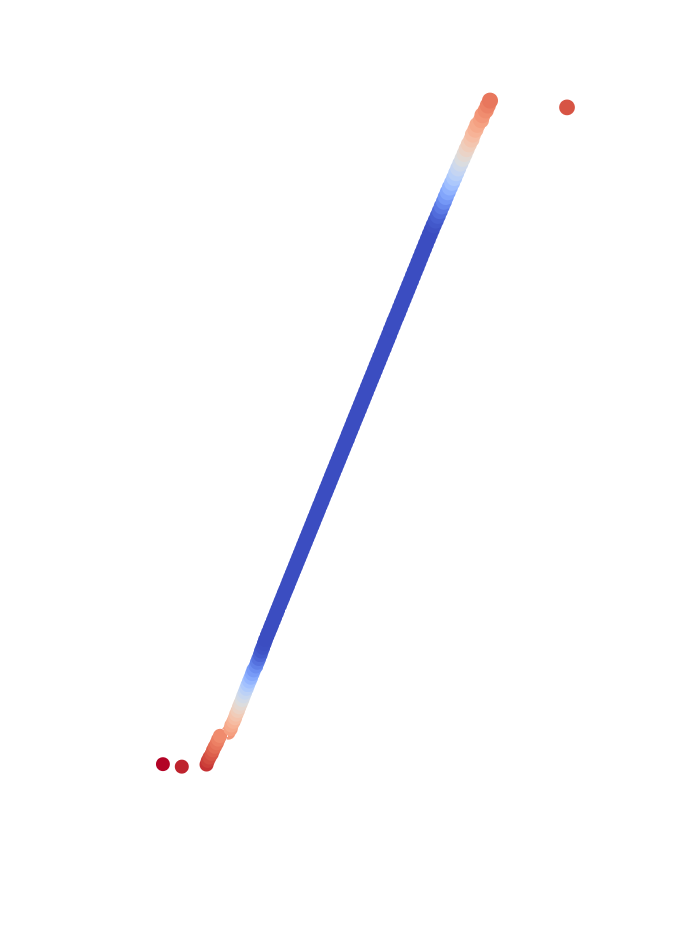}} & &
				\centered{\includegraphics[width=0.16\linewidth]{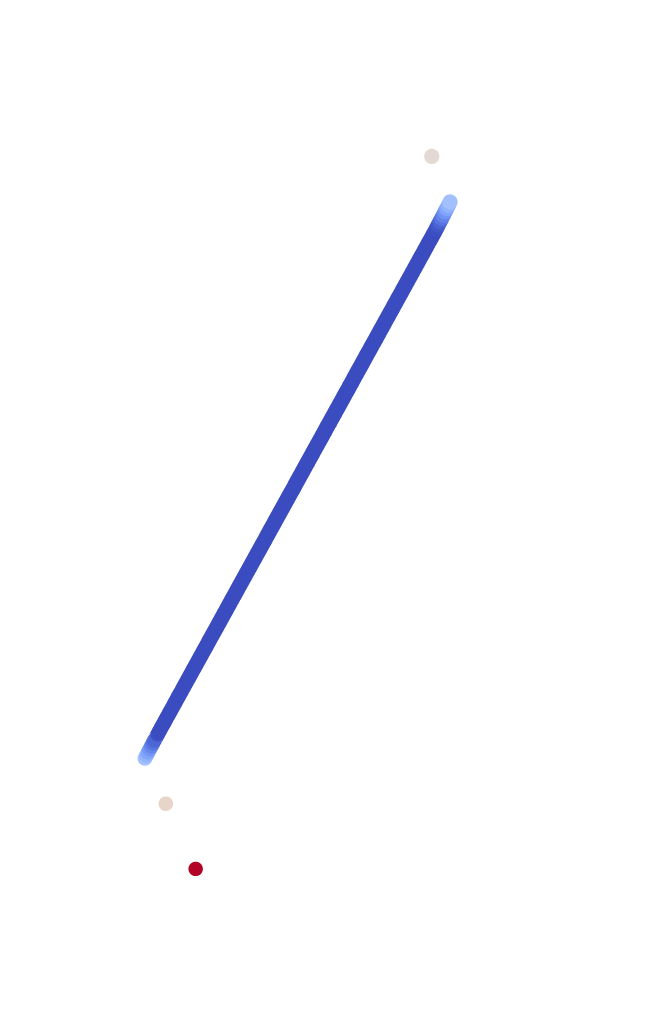}}\\
				\rotatebox[origin=c]{90}{Channel Chanalyzer} &  &
				\centered{\includegraphics[width=0.19\linewidth]{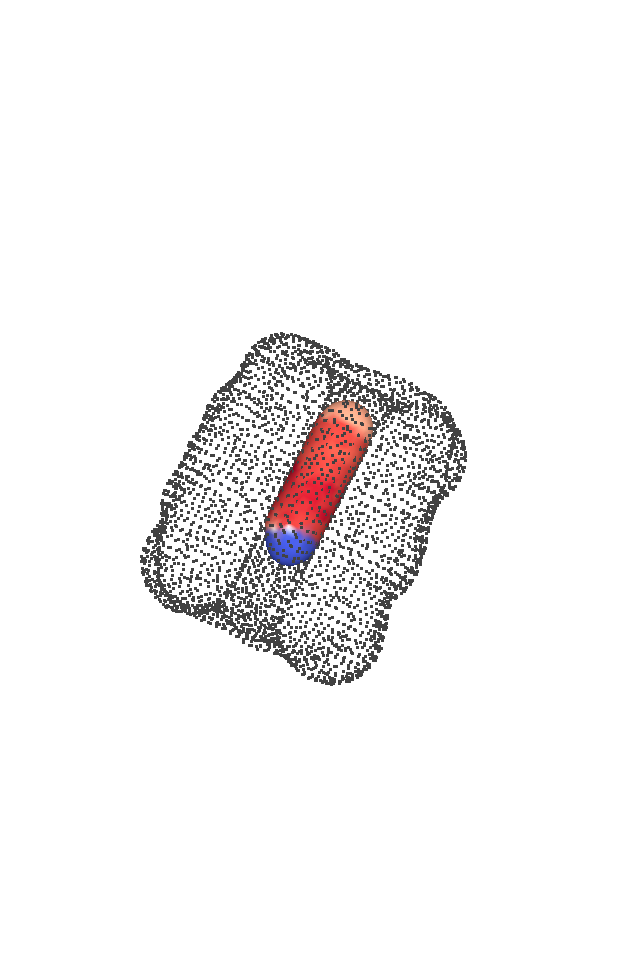}} & &\centered{\includegraphics[width=0.19\linewidth]{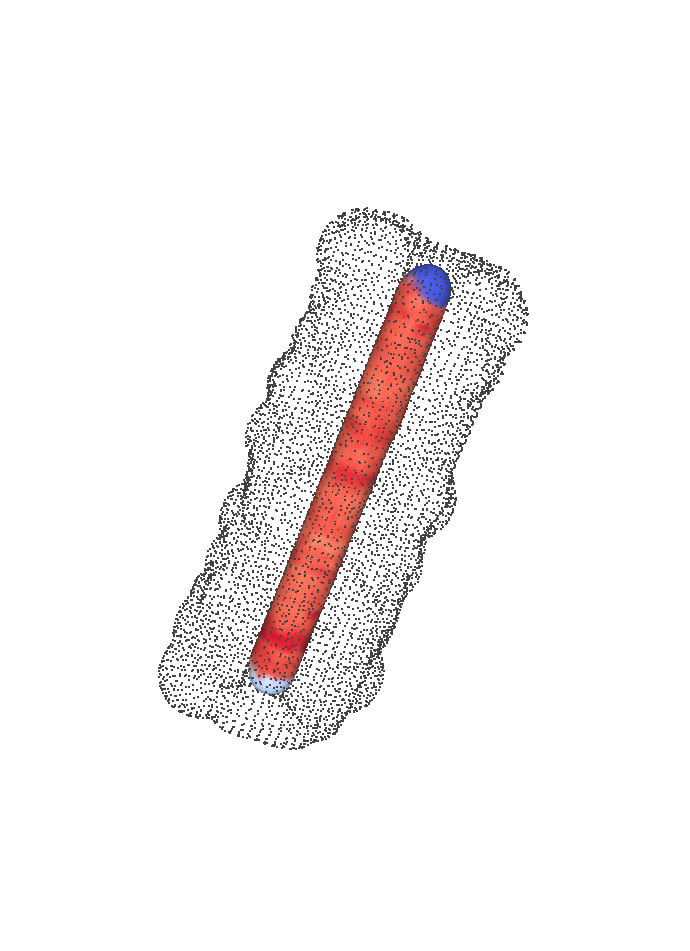}} & &
				\centered{\includegraphics[width=0.19\linewidth]{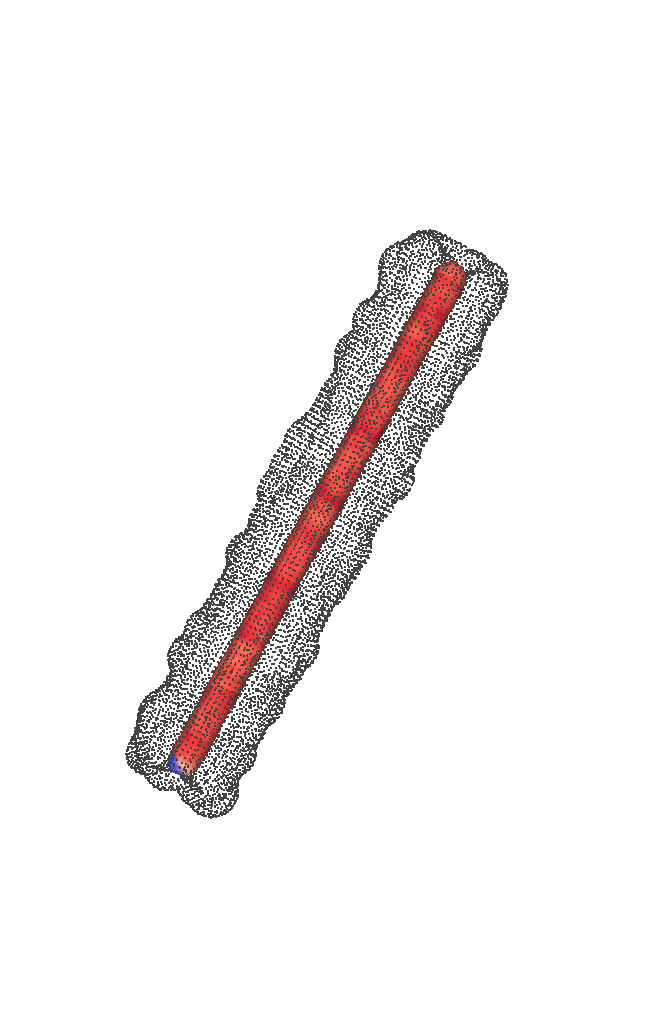}}\\
				\rotatebox[origin=c]{90}{Channel 1 HOLE} &  &
				\centered{\includegraphics[width=0.19\linewidth]{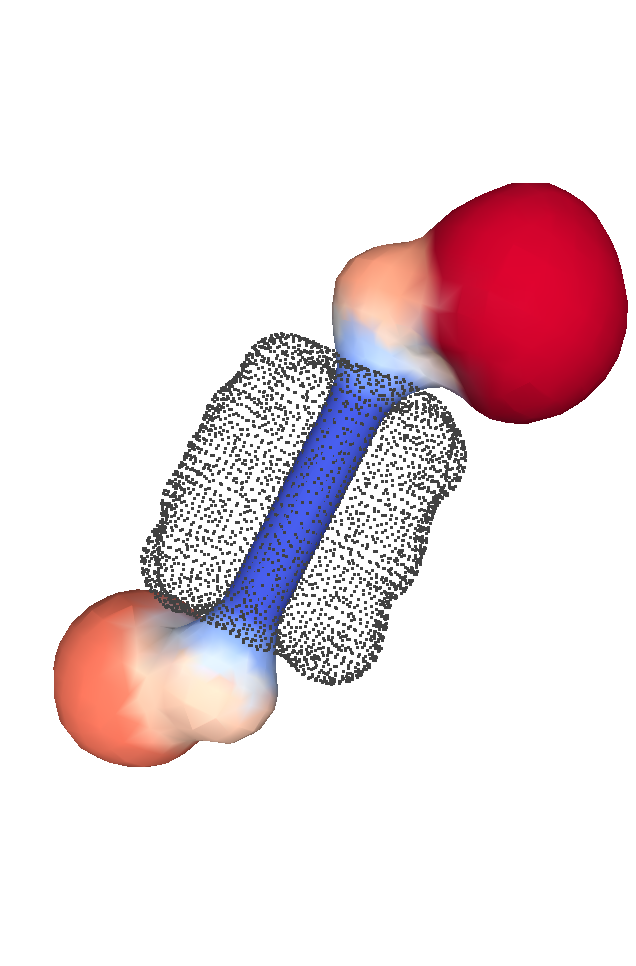}} & &\centered{\includegraphics[width=0.19\linewidth]{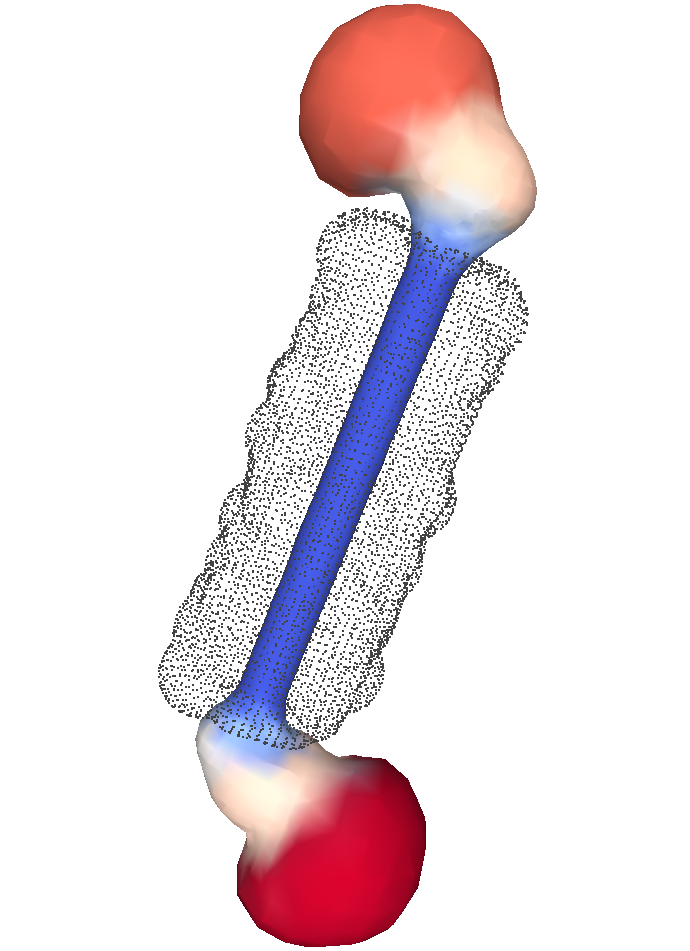}} & &
				\centered{\includegraphics[width=0.19\linewidth]{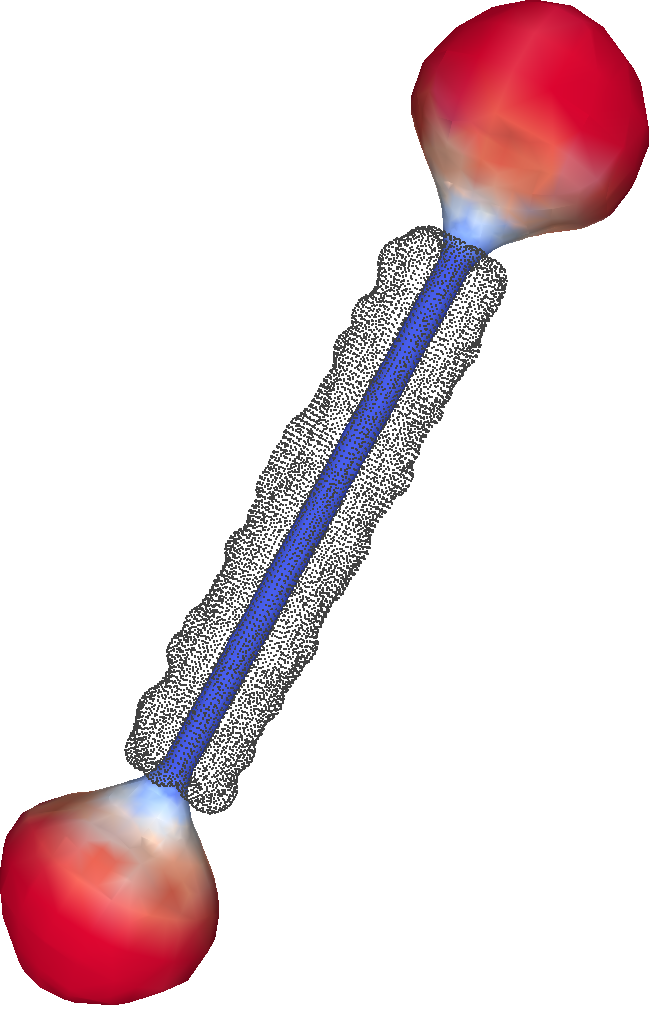}}\\
				\rotatebox[origin=c]{90}{Channel 2 HOLE} &  &
				\centered{\includegraphics[width=0.19\linewidth]{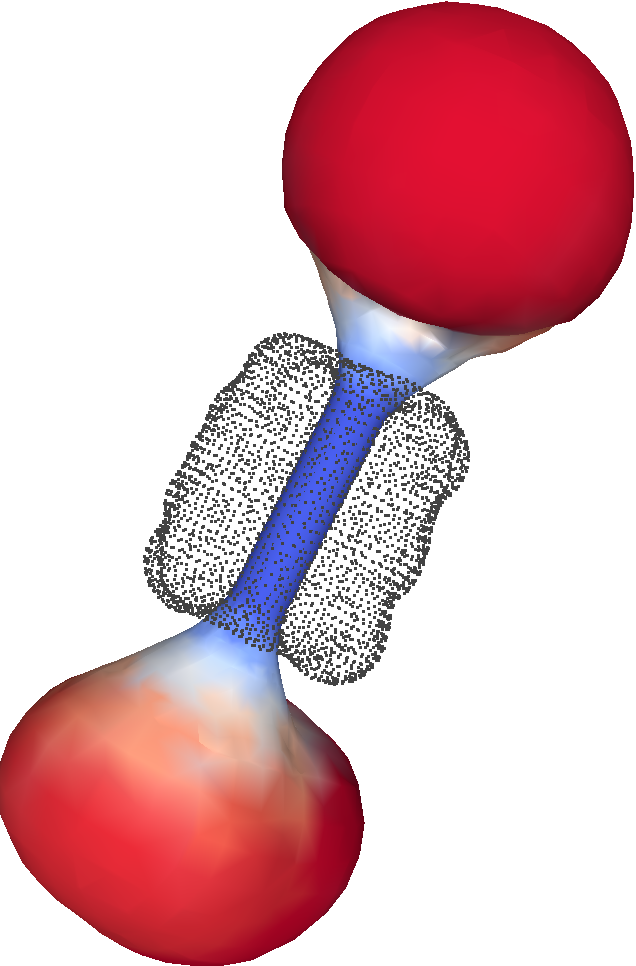}} & &\centered{\includegraphics[width=0.19\linewidth]{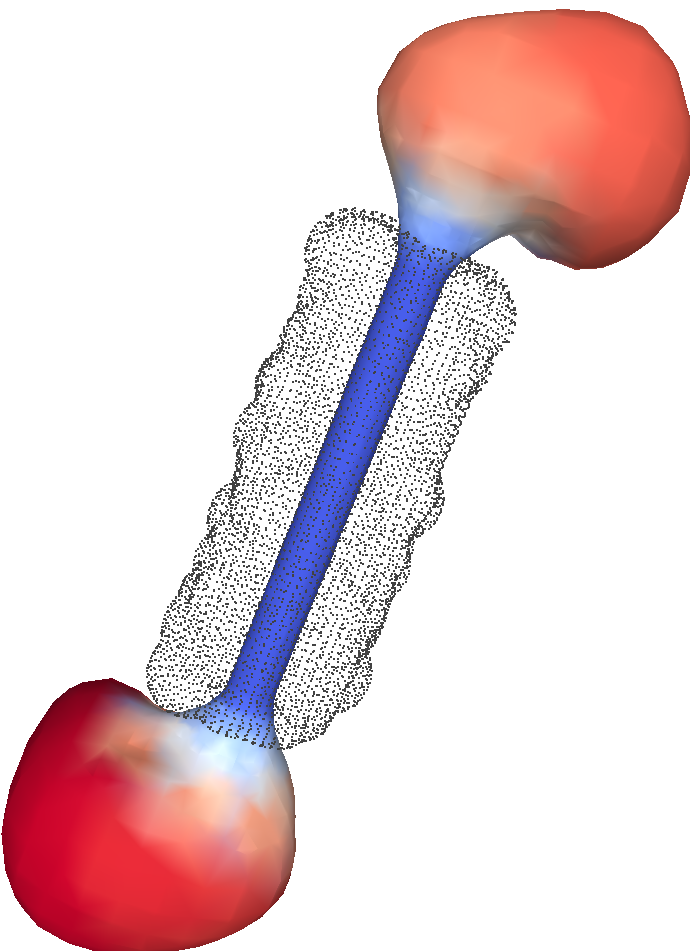}} & &
				\centered{\includegraphics[width=0.19\linewidth]{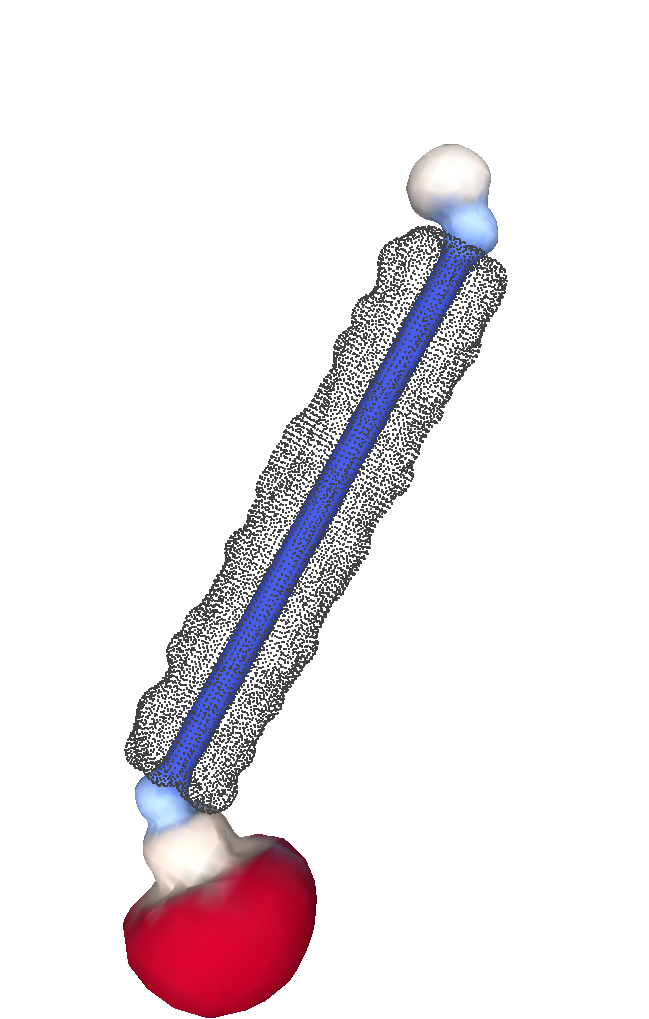}}\\
			\end{tabular}
		}
	\end{center}
	\caption{Centerlines and channels obtained by running  Chanalyzer and HOLE (w.r.t. two different initial positions) on the benchmarked models. Channels are represented as a collection of spheres centred at the points $\mathbf{p}$ of the centerline and of radius $\rho_\mathbf{p}$. Centerlines and sphere points are coloured in accordance with the radius values $\rho_\mathbf{p}$ of the point $\mathbf{p}$ which are in correspondence to. In the pictures, the Solvent Excluded Surface is depicted in grey. The colour scheme adopted for representing radius value is the ``coolwarm'' colourmap in Matplotlib \cite{Hunter:2007}. It ranges from dark blue for low values to dark red for high values passing from white. In order to improve the visualization as well as the understanding, we decided to normalize the scale to each of the considered examples. The minimum and the maximum values expressed in $\AA$ are reported in Table \ref{tab:radius_fake}.}
	\label{fig:comparison_fake}
\end{figure*}

\begin{figure*}[h!]
	\begin{center}
		\resizebox{0.99\textwidth }{!}{
			\begin{tabular}{c  c  c} 
				BM 1 & BM 2 & BM 3 \\
				\includegraphics[width=0.3\linewidth]{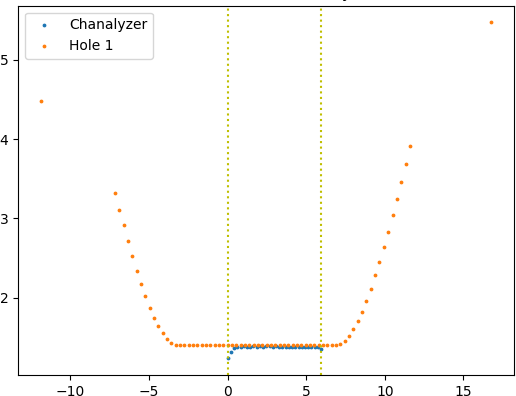} &
				\includegraphics[width=0.3\linewidth]{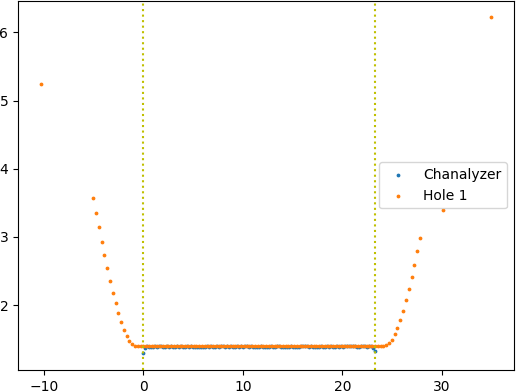} &
				\includegraphics[width=0.3\linewidth]{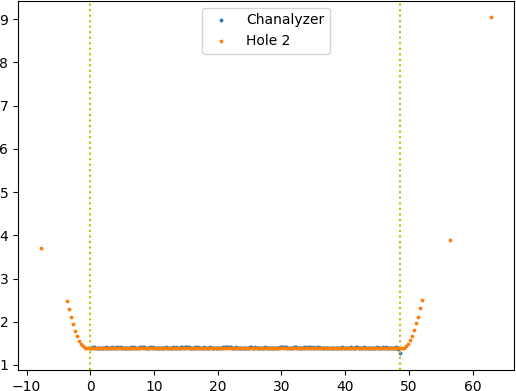}\\
				
			\end{tabular}
		}
	\end{center}
	\caption{Graphs of the radius functions of the centerlines of the models considered in Figure \ref{fig:comparison_fake}. For each structure, the centerline retrieved by Chanalyzer is depicted in blue while we represent in orange the one among the two produced by HOLE obtaining the higher matching score. Moreover, vertical dashed lines denote the extrema of the interval in which the two centerlines are identified as matched.}
	\label{fig:radius_fake}
\end{figure*}

First, we run the two tools on the three benchmarked models introduced in Section \ref{dataset} on which we can make quantitative estimates with respect to ground truth. Figure \ref{fig:comparison_fake} visually depicts the results obtained by running Chanalyzer and HOLE (with respect to two different initial positions).
Table \ref{tab:radius_fake} reports the average, minimum, and maximum values of the radius of the retrieved channels obtained by running Chanalyzer and HOLE on the benchmarked models. 
In consideration of the fact that for the benchmarked models the channel radius (i.e., the radius of the maximal inscribed ball) has been set to $1.4\AA$, we can observe that Chanalyzer achieves accurate average results and a small difference between the minimum and the maximum values. Differently, 
as shown in Figure \ref{fig:comparison_fake}, the channels returned by HOLE protrude outward from the model surface. This affects the obtained radius values. As confirmed by Figure \ref{fig:radius_fake}, the radius values achieved by HOLE are correct for the central portion of the retrieved channel while they significantly grow at the channel extremities.  
A complete quantitative validation is provided in Table \ref{tab:single_analysis_fake}. It is clear that Chanalyzer returns shorter centerlines and lower volumes but with higher straightness parameters. Table \ref{tab:comparison_fake} compares centerlines and radii estimated by the two tools in each of the three you models: from the numbers within we can conclude that centerlines provided by Chanalyzer correspond to a subpart of those extracted by HOLE, with just small differences in the radius values. A similar conclusion is reached from Figure \ref{fig:radius_fake}.

\begin{table*}[h!]\centering
	\caption{Radius values of the channels obtained by running Chanalyzer (Chan.) and HOLE (H.1 and H.2, with respect to two different initial positions) on the benchmarked models. Specifically, we report the average, minimum, and maximum values of the radius of the correspondent channel. Values are expressed in $\AA$.}
	\label{tab:radius_fake}
	\begin{tabular}{|c|ccc|ccc|ccc|}
		\hline 
		\cellcolor{ForestGreenOriginal!35} &  \multicolumn{3}{c|}{\cellcolor{ForestGreenOriginal!35} Average radius} & \multicolumn{3}{c|}{\cellcolor{ForestGreenOriginal!35} Min radius} & \multicolumn{3}{c|}{ \cellcolor{ForestGreenOriginal!35} Max radius} \\
		\cellcolor{ForestGreenOriginal!35} Model & \cellcolor{ForestGreenOriginal!35} Chan. & \cellcolor{ForestGreenOriginal!35} H.1 & \cellcolor{ForestGreenOriginal!35} H.2 & \cellcolor{ForestGreenOriginal!35} Chan. & \cellcolor{ForestGreenOriginal!35} H.1 & \cellcolor{ForestGreenOriginal!35} H.2 & \cellcolor{ForestGreenOriginal!35} Chan. & \cellcolor{ForestGreenOriginal!35} H.1 & \cellcolor{ForestGreenOriginal!35} H.2 \\
		\cline{1-10}
		
		BM 1 & 1.37 & 1.90 & 3.17 & 1.23 & 1.40 & 1.39 & 1.39 & 5.47 & 8.22 \\
		BM 2 & 1.39 & 1.67 & 2.35 & 1.29 & 1.40 & 1.39 & 1.40 & 6.22 & 8.05 \\ 
		BM 3 & 1.40 & 2.53 & 1.51 & 1.27 & 1.40 & 1.40 & 1.41 & 9.93 & 9.04  \\ 
		\hline
	\end{tabular}
\end{table*}

\begin{table*}[h!]\centering
	\caption{Measure values of the channels obtained by running Chanalyzer (Chan.) and HOLE (H.1 and H.2, with respect to two different initial positions) on the benchmarked models. Specifically, we report the number of points of each centerline, their length, their straightness, and the volume of the correspondent channel (represented as the region contained in a collection of spheres).}
	\label{tab:single_analysis_fake}
	\resizebox{\columnwidth}{!}{
	\begin{tabular}{|c|ccc|ccc|ccc|ccc|}
		\hline 
		\cellcolor{ForestGreenOriginal!35} &  \multicolumn{3}{c|}{\cellcolor{ForestGreenOriginal!35} $\#$ points} & \multicolumn{3}{c|}{\cellcolor{ForestGreenOriginal!35} Length} & \multicolumn{3}{c|}{ \cellcolor{ForestGreenOriginal!35} Straightness} & \multicolumn{3}{c|}{ \cellcolor{ForestGreenOriginal!35} Volume}\\
		\cellcolor{ForestGreenOriginal!35} Model & \cellcolor{ForestGreenOriginal!35} Chan. & \cellcolor{ForestGreenOriginal!35} H.1 & \cellcolor{ForestGreenOriginal!35} H.2 & \cellcolor{ForestGreenOriginal!35} Chan. & \cellcolor{ForestGreenOriginal!35} H.1 & \cellcolor{ForestGreenOriginal!35} H.2 & \cellcolor{ForestGreenOriginal!35} Chan. & \cellcolor{ForestGreenOriginal!35} H.1 & \cellcolor{ForestGreenOriginal!35} H.2 & \cellcolor{ForestGreenOriginal!35} Chan. & \cellcolor{ForestGreenOriginal!35} H.1 & \cellcolor{ForestGreenOriginal!35} H.2 \\
		\cline{1-13}
		
		BM 1 & 30 & 71 & 113 & 5.94 & 28.68 & 36.66 & 4.68 & 0.23 & 0.16 & 44.32 & 1228.08 & 4986.24 \\
		BM 2 & 117 & 121 & 162 & 23.28 & 45.28 & 48.01 & 2.80 & 0.15 & 0.14 & 150.58 & 1736.37 & 3877.39 \\ 
		BM 3 & 243 & 290 & 187 & 48.56 & 72.58 & 70.79 & 1.38 & 1.35 & 0.12 & 310.63 & 8657.38 & 3170.73 \\ 
		\hline
	\end{tabular}
}
\end{table*}

\begin{table*}[h!]\centering
	\caption{Comparison measure values between the channels obtained by running Chanalyzer (Ch.) and HOLE (H.1 and H.2, with respect to two different initial positions) on the benchmarked models. Specifically, we report in columns $match$ as the percentage of points of a centerline matched with a different centerline. In columns $d_\rho$, we collect the distance between the radius functions of two centerlines on their portion identified as matched.}
	\label{tab:comparison_fake}
	\begin{tabular}{|c|cccc|cc|}
		\hline 
		\cellcolor{ForestGreenOriginal!35} &  \multicolumn{4}{c|}{\cellcolor{ForestGreenOriginal!35} $match$} & \multicolumn{2}{c|}{\cellcolor{ForestGreenOriginal!35} $d_\rho$} \\
		\cellcolor{ForestGreenOriginal!35} Models & \cellcolor{ForestGreenOriginal!35} (Ch., H.1) & \cellcolor{ForestGreenOriginal!35} (H.1, Ch.) & \cellcolor{ForestGreenOriginal!35} (Ch., H.2) & \cellcolor{ForestGreenOriginal!35} (H.2, Ch.) & \cellcolor{ForestGreenOriginal!35} (Ch., H.1) & \cellcolor{ForestGreenOriginal!35} (Ch., H.2) \\
		\cline{1-7}
		
		BM 1 & 100.00 & 40.85 & 100.00 & 27.43 & 0.05 & 0.05 \\
		BM 2 & 100.00 & 74.38 & 100.00 & 60.49 & 0.03 & 0.03 \\ 
		BM 3 & 100.00 & 69.31 & 100.00 & 88.77 & 0.02 & 0.02 \\ 
		\hline
	\end{tabular}
\end{table*}

\begin{figure*}[h!]
	\begin{center}
		\resizebox{0.75\textwidth }{!}{	
			\begin{tabular}{c  c  c  c  c } 
				& 5EK0 & 7K48 & 7W9K & 8FHD \\
				& & & & \\
				\rotatebox[origin=c]{90}{Centerline Chanalyzer}&
				\centered{\includegraphics[width=0.16\linewidth]{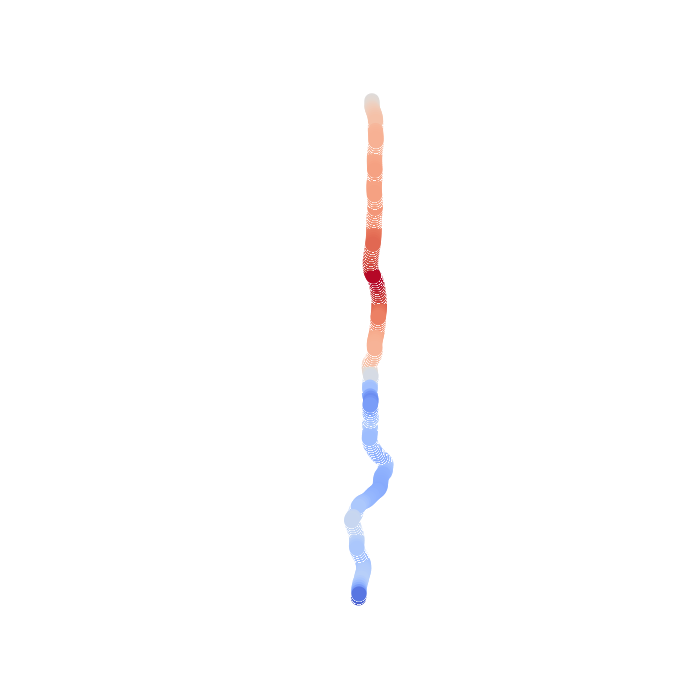} }& \centered{\includegraphics[width=0.16\linewidth]{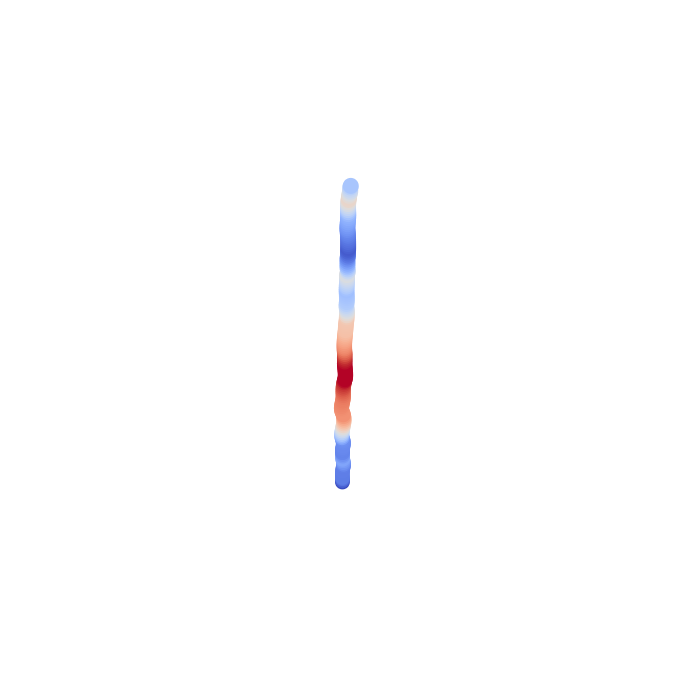}} &
				\centered{\includegraphics[width=0.16\linewidth]{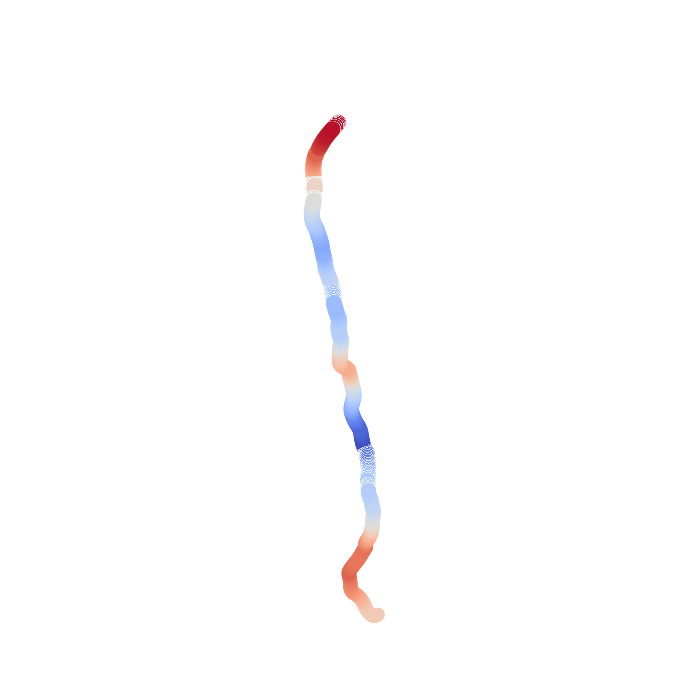}} &
				\centered{\includegraphics[width=0.16\linewidth]{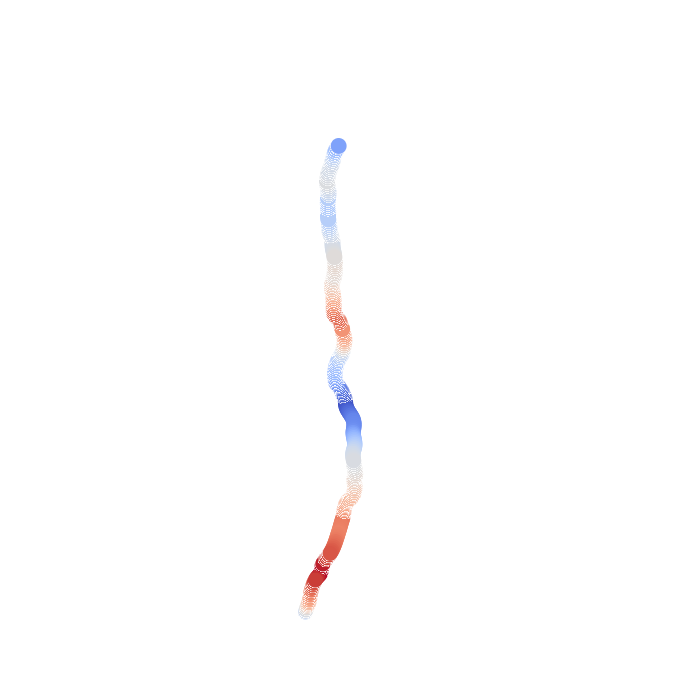}}\\
				& & & & \\
				\rotatebox[origin=c]{90}{Centerline 1 HOLE} &
				\centered{\includegraphics[width=0.16\linewidth]{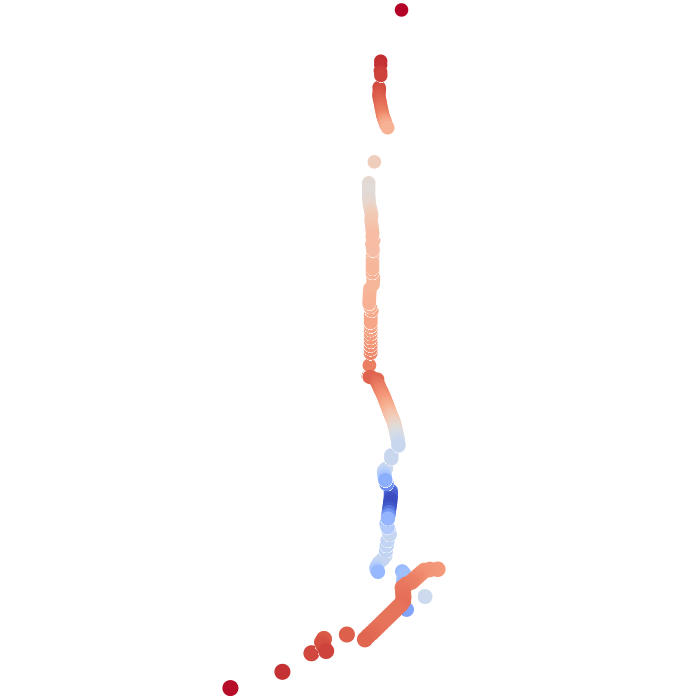}} & \centered{\includegraphics[width=0.16\linewidth]{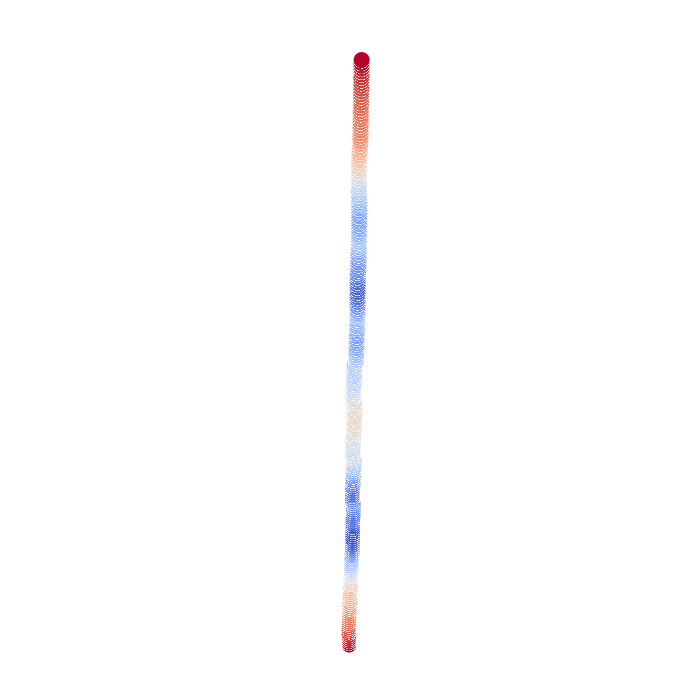}} &
				\centered{\includegraphics[width=0.16\linewidth]{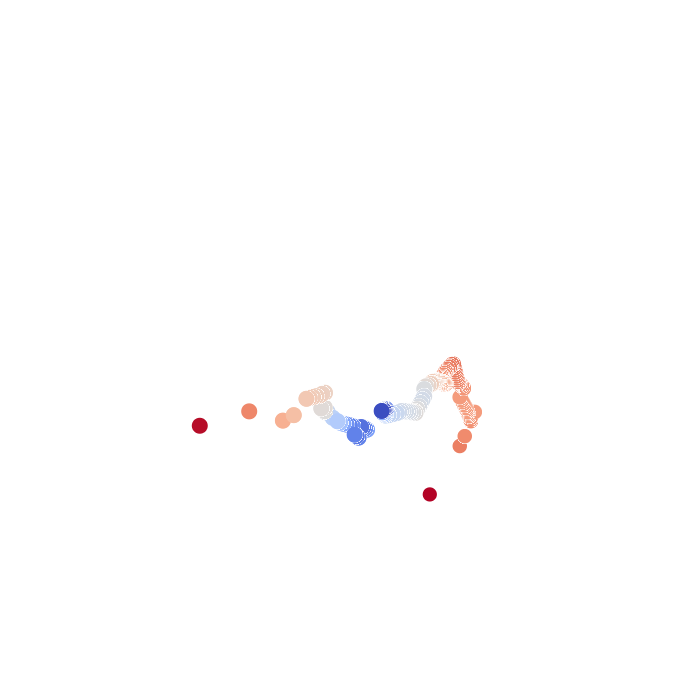}} &
				\centered{\includegraphics[width=0.16\linewidth]{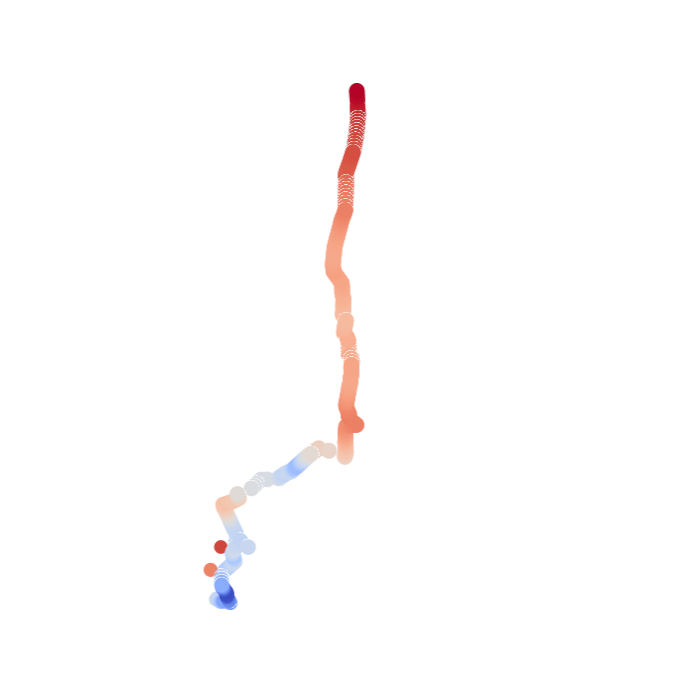}}\\
				& & & & \\
				\rotatebox[origin=c]{90}{Centerline 2 HOLE} &
				\centered{\includegraphics[width=0.16\linewidth]{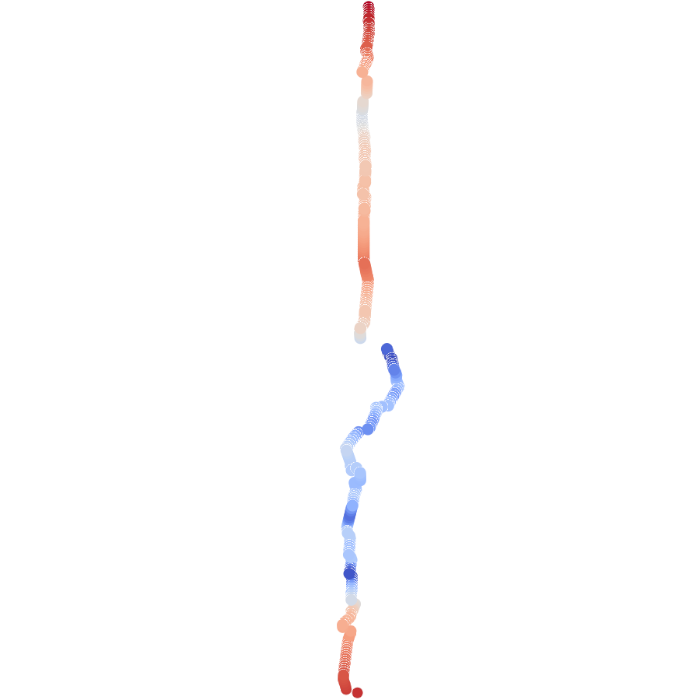}} & \centered{\includegraphics[width=0.16\linewidth]{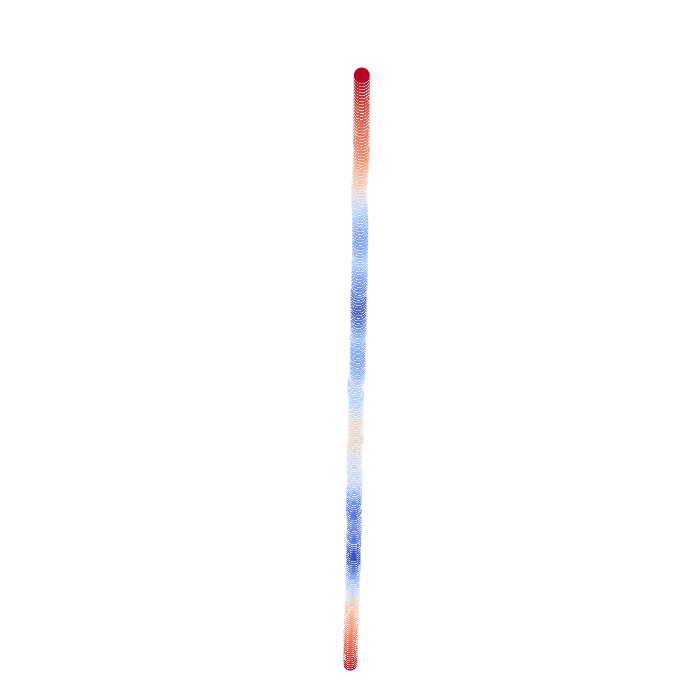}} &
				\centered{\includegraphics[width=0.16\linewidth]{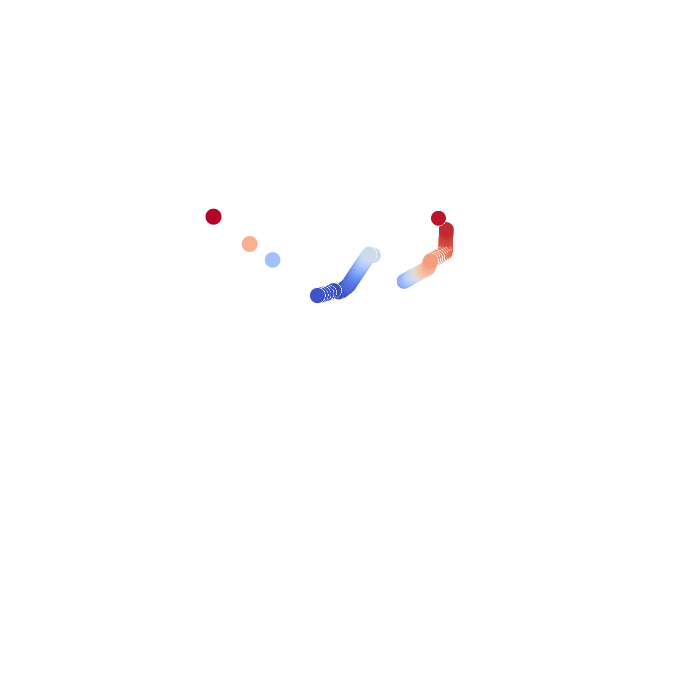}} &
				\centered{\includegraphics[width=0.16\linewidth]{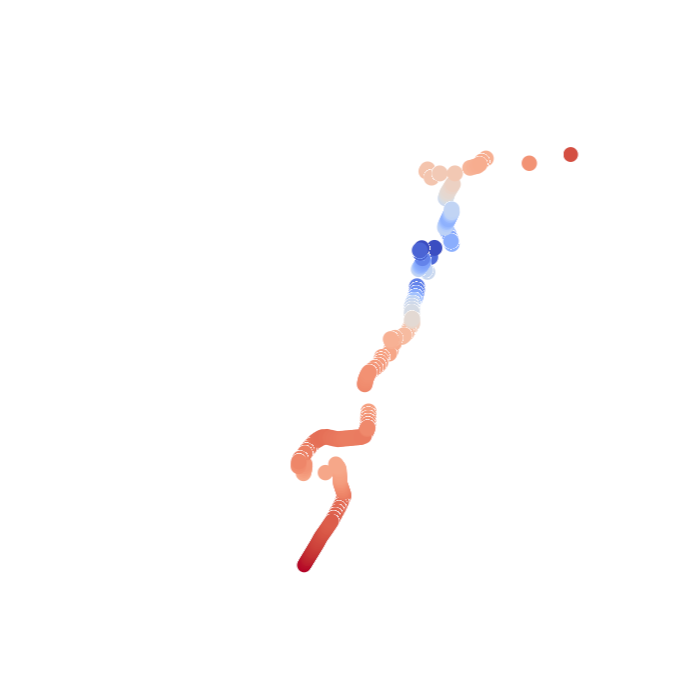}}\\
				& & & & \\
				\rotatebox[origin=c]{90}{Channel Chanalyzer} &
				\centered{\includegraphics[width=0.19\linewidth]{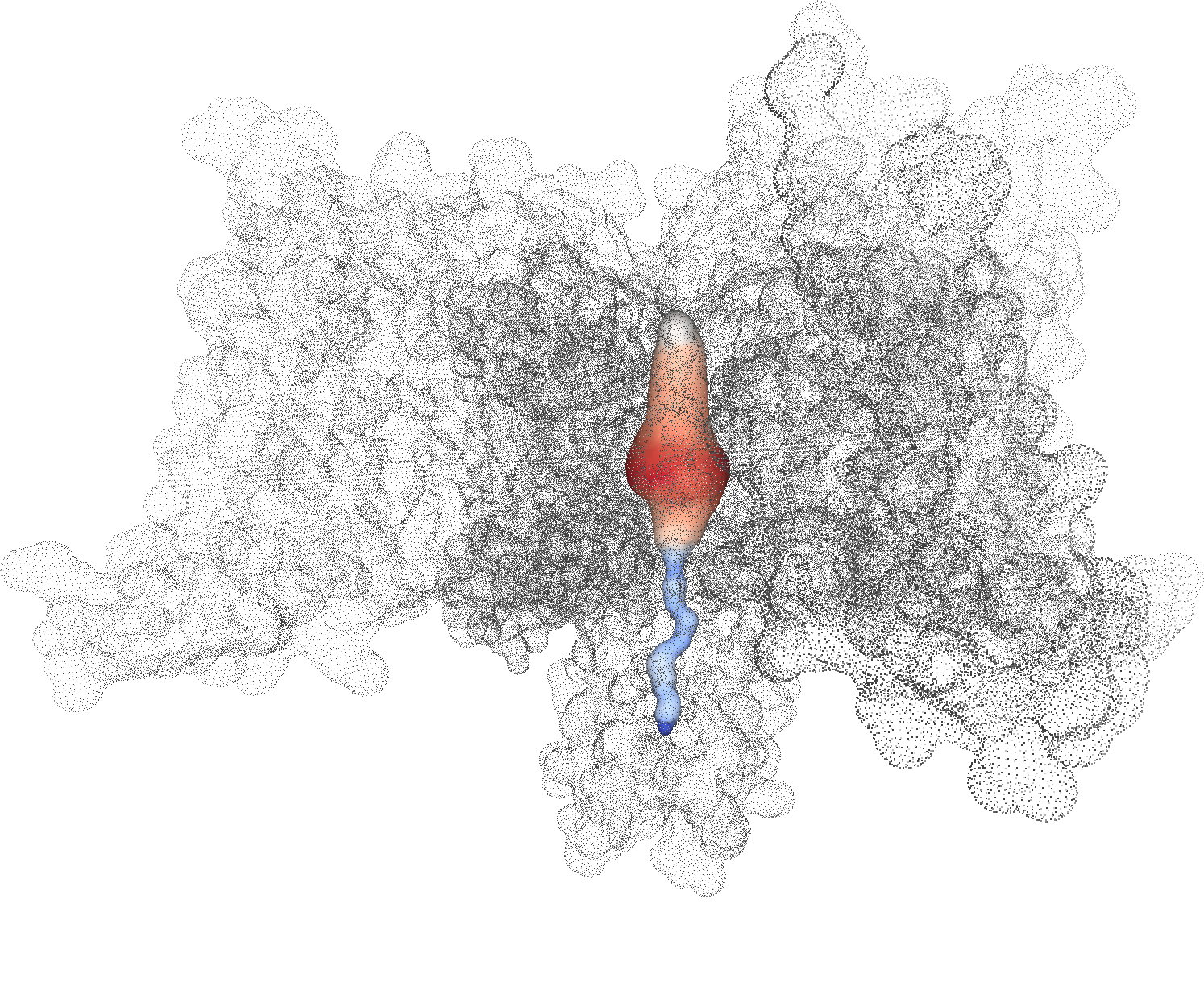}} & \centered{\includegraphics[width=0.19\linewidth]{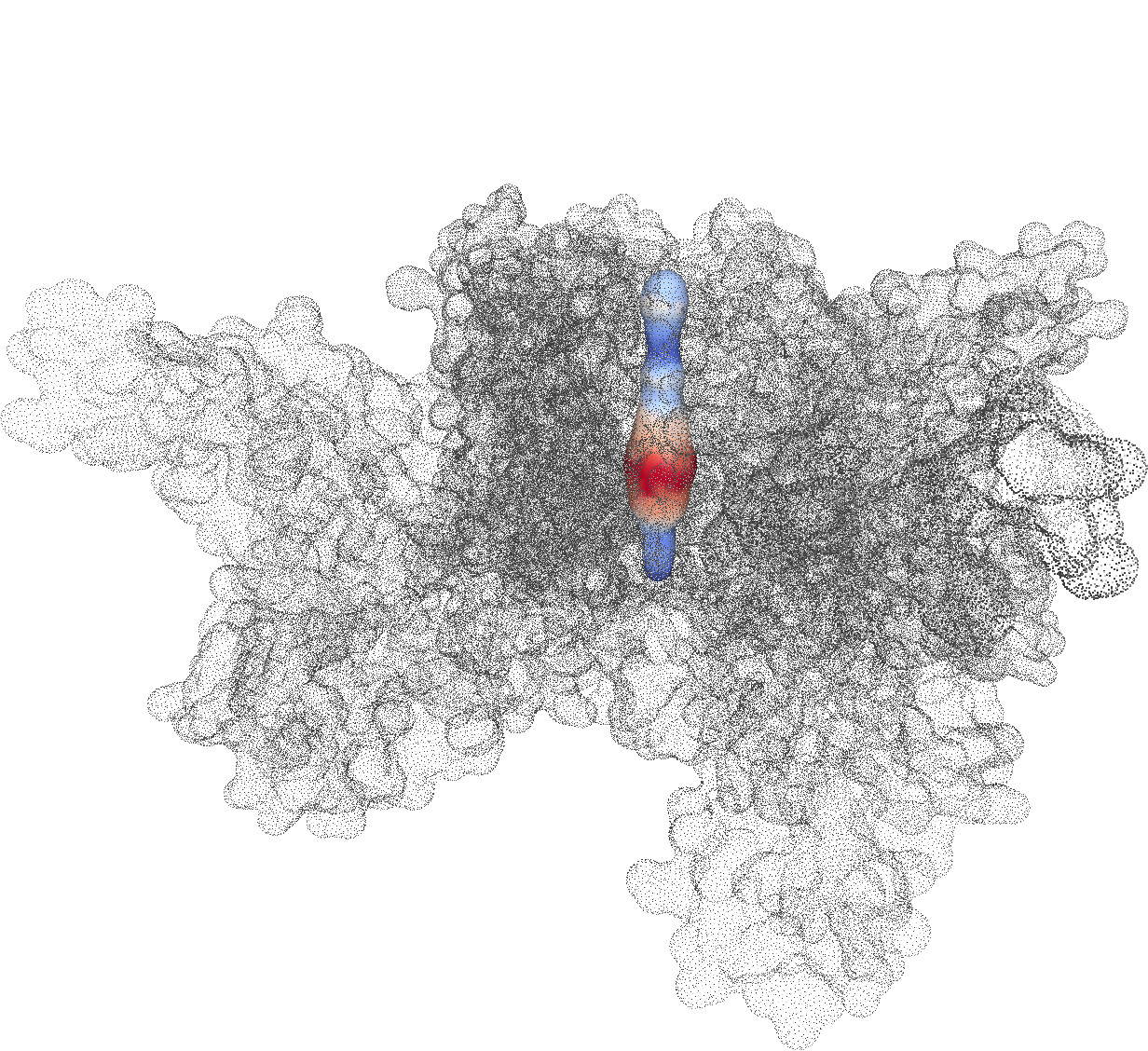}} &
				\centered{\includegraphics[width=0.19\linewidth]{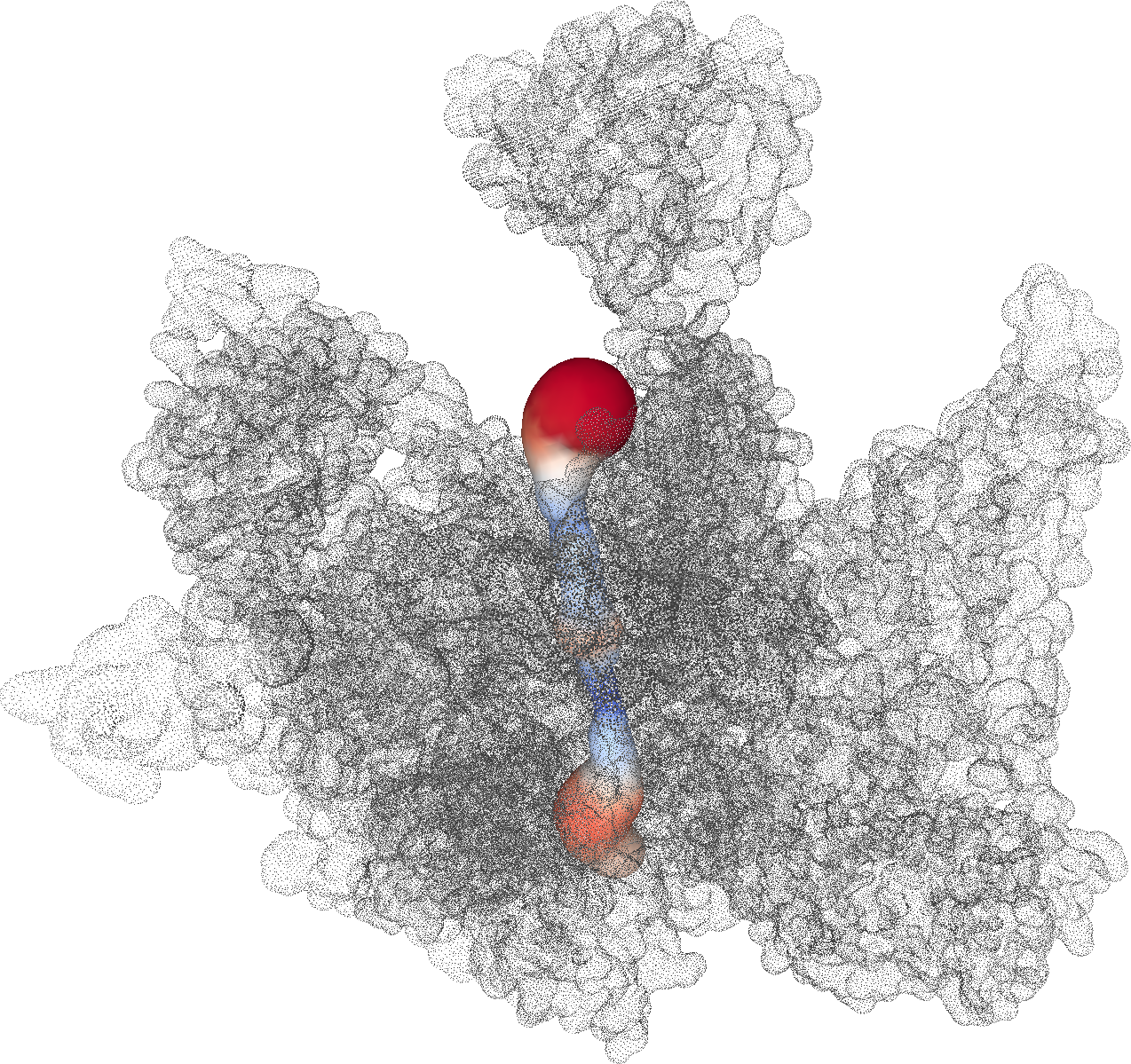}} &
				\centered{\includegraphics[width=0.19\linewidth]{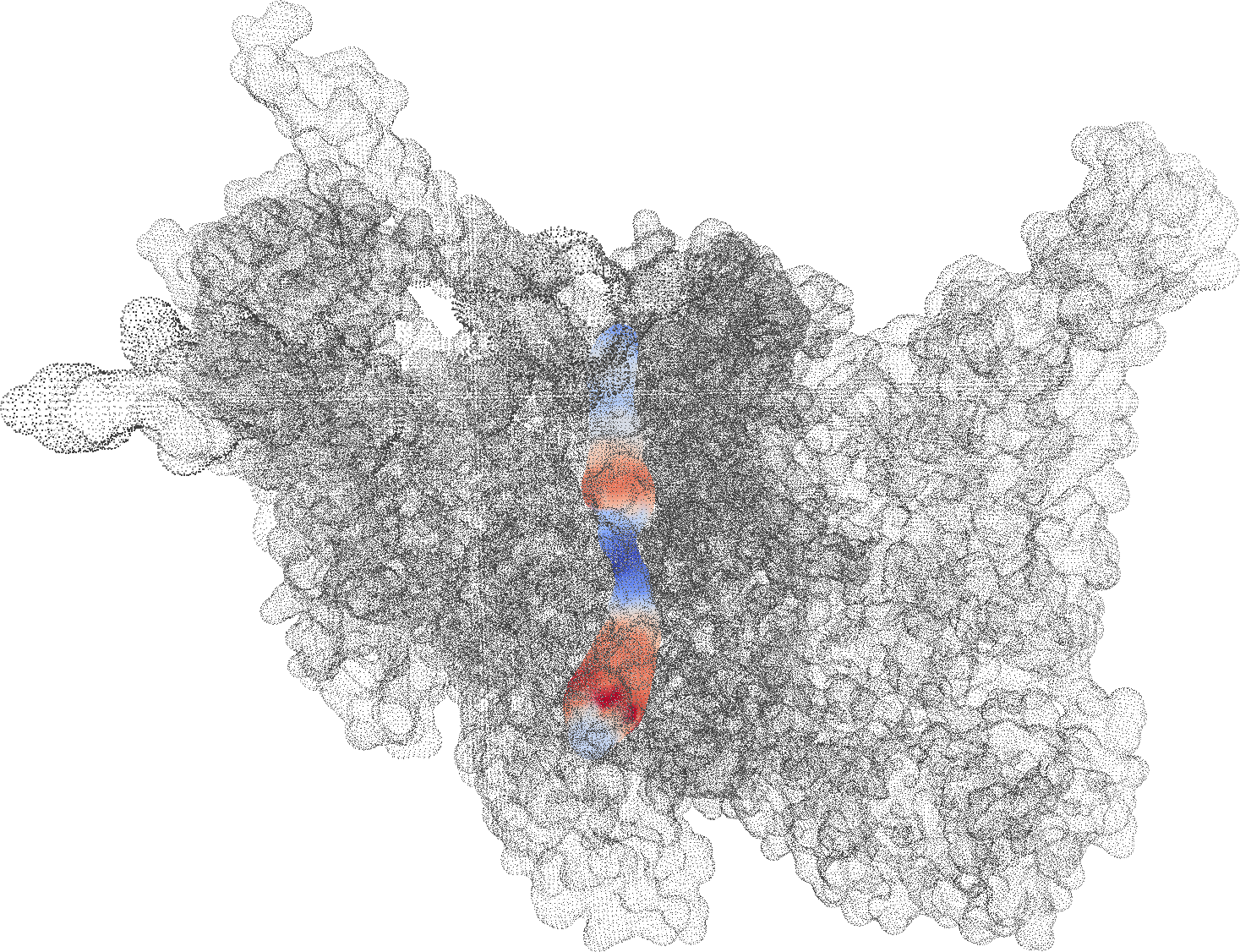}}\\
				& & & & \\
				\rotatebox[origin=c]{90}{Channel 1 HOLE} &
				\centered{\includegraphics[width=0.19\linewidth]{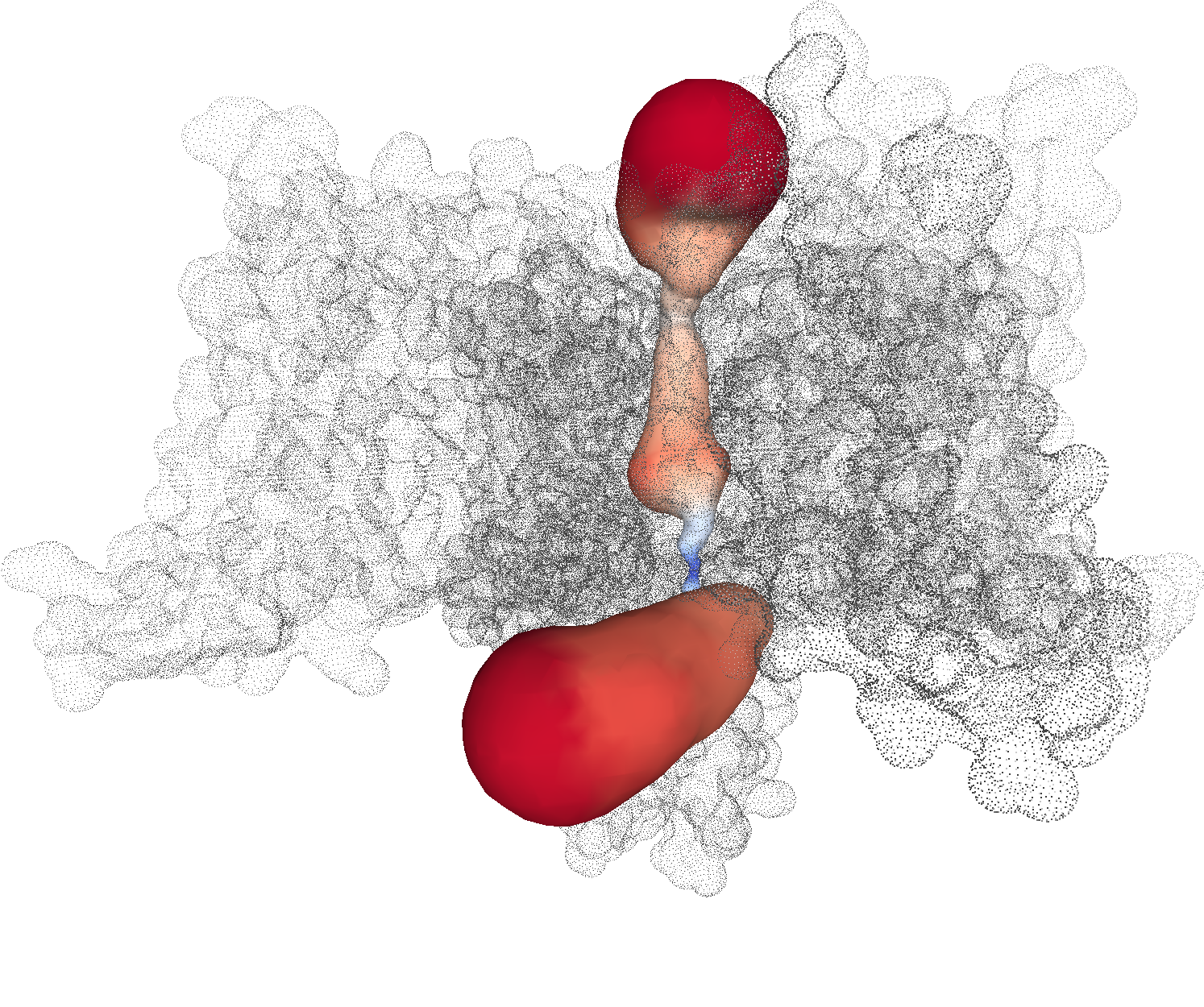}} & \centered{\includegraphics[width=0.19\linewidth]{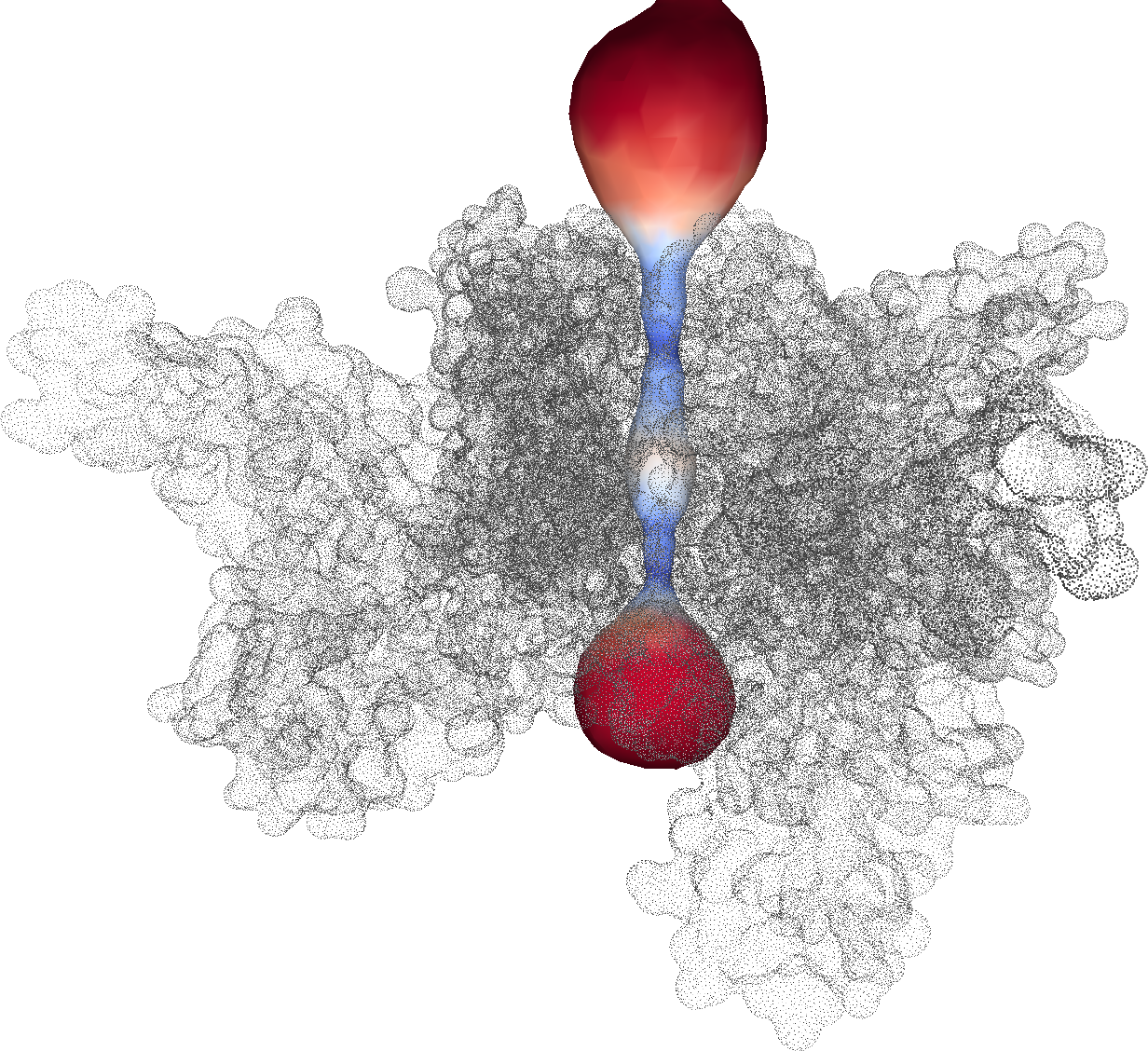}} &
				\centered{\includegraphics[width=0.19\linewidth]{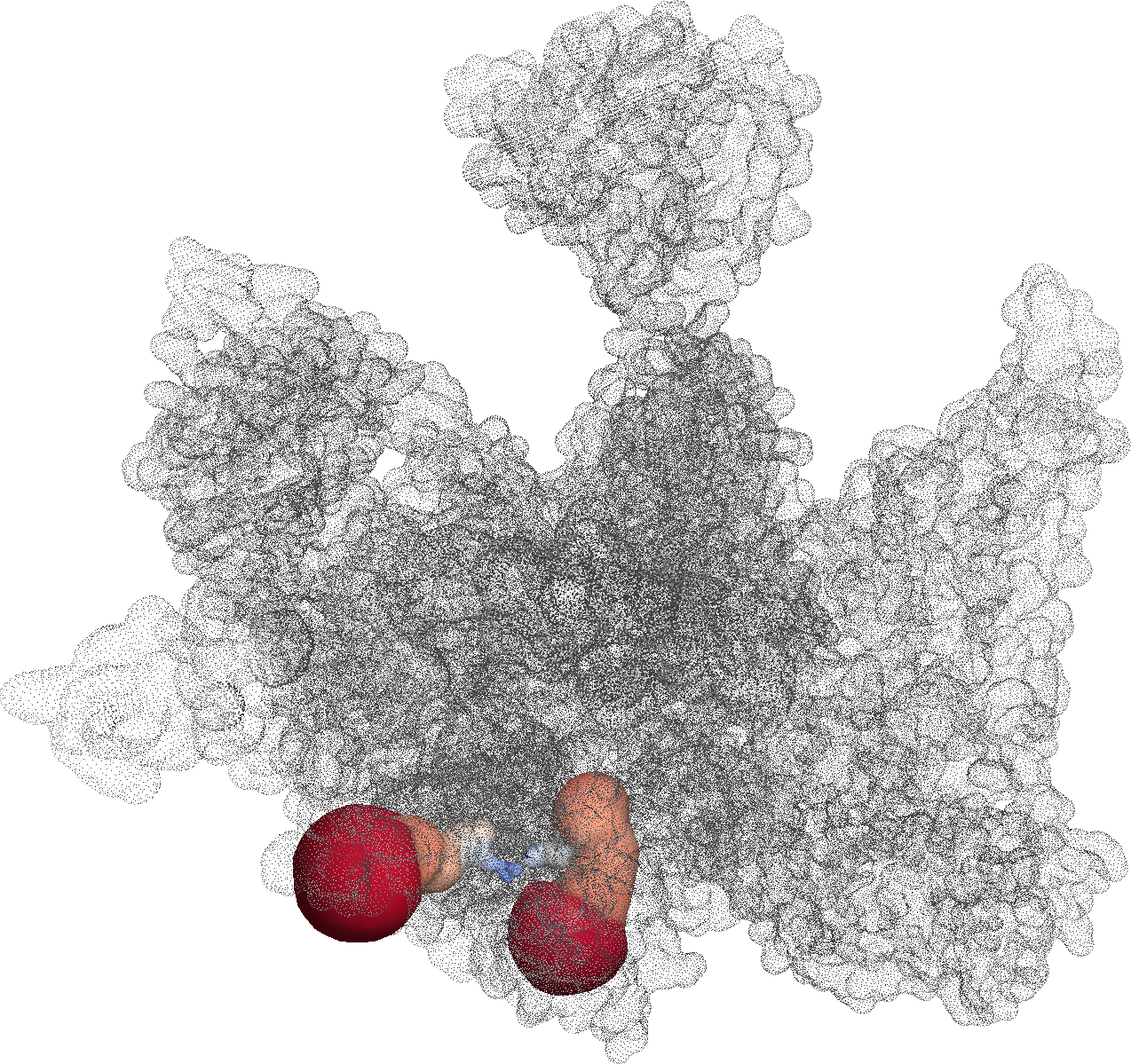}} &
				\centered{\includegraphics[width=0.19\linewidth]{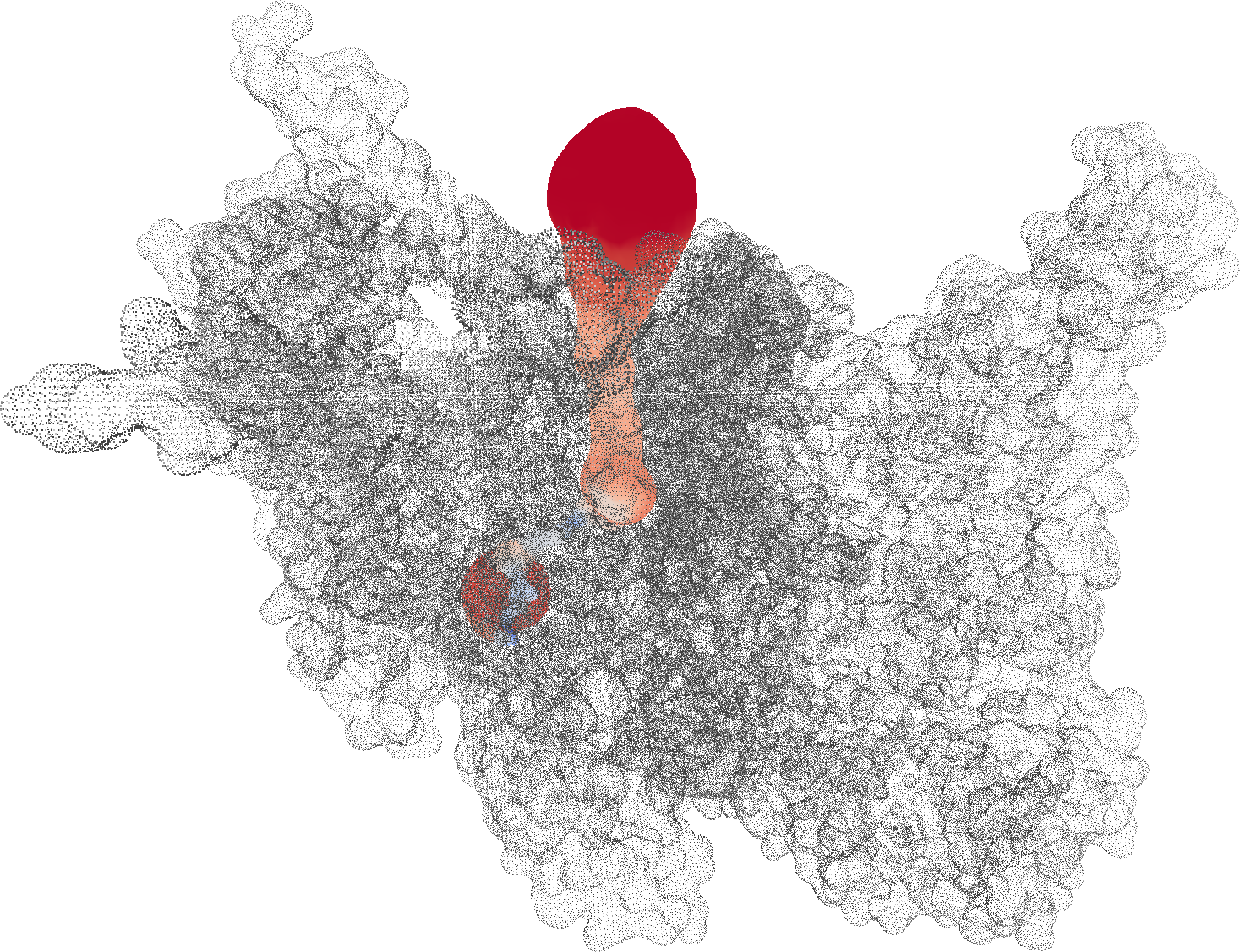}}\\
				& & & & \\
				\rotatebox[origin=c]{90}{Channel 2 HOLE} &
				\centered{\includegraphics[width=0.19\linewidth]{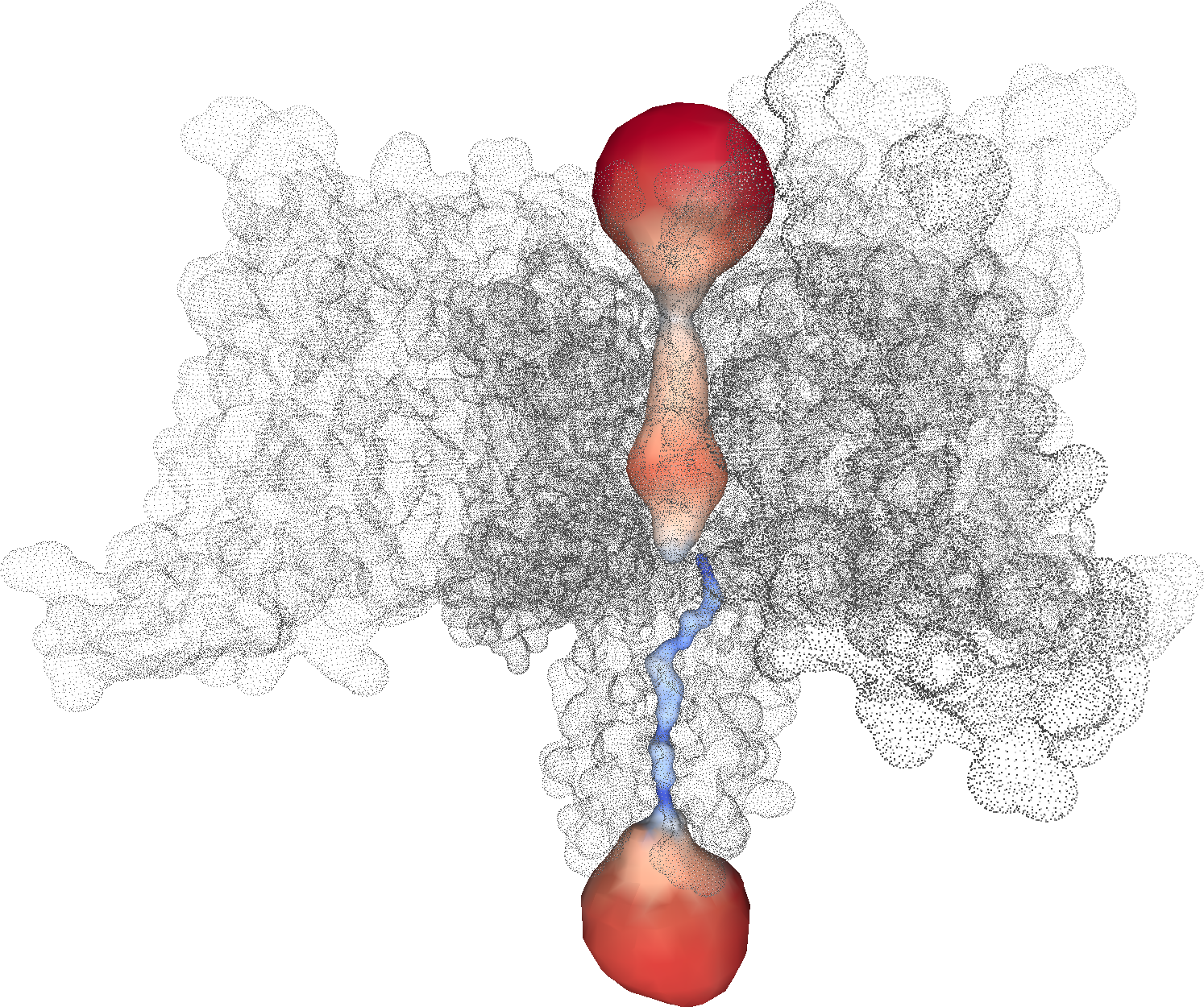}} & \centered{\includegraphics[width=0.19\linewidth]{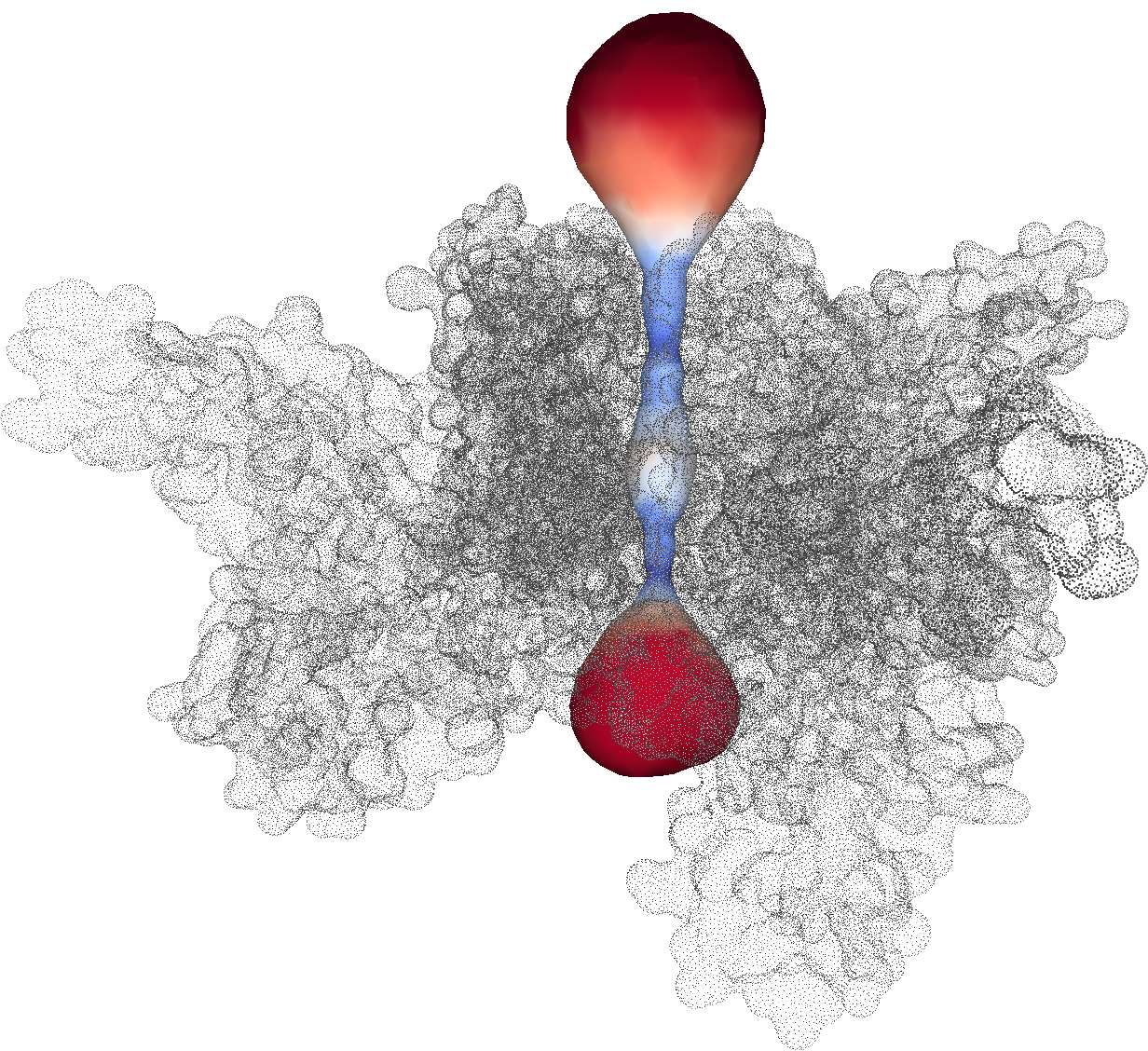}} &
				\centered{\includegraphics[width=0.19\linewidth]{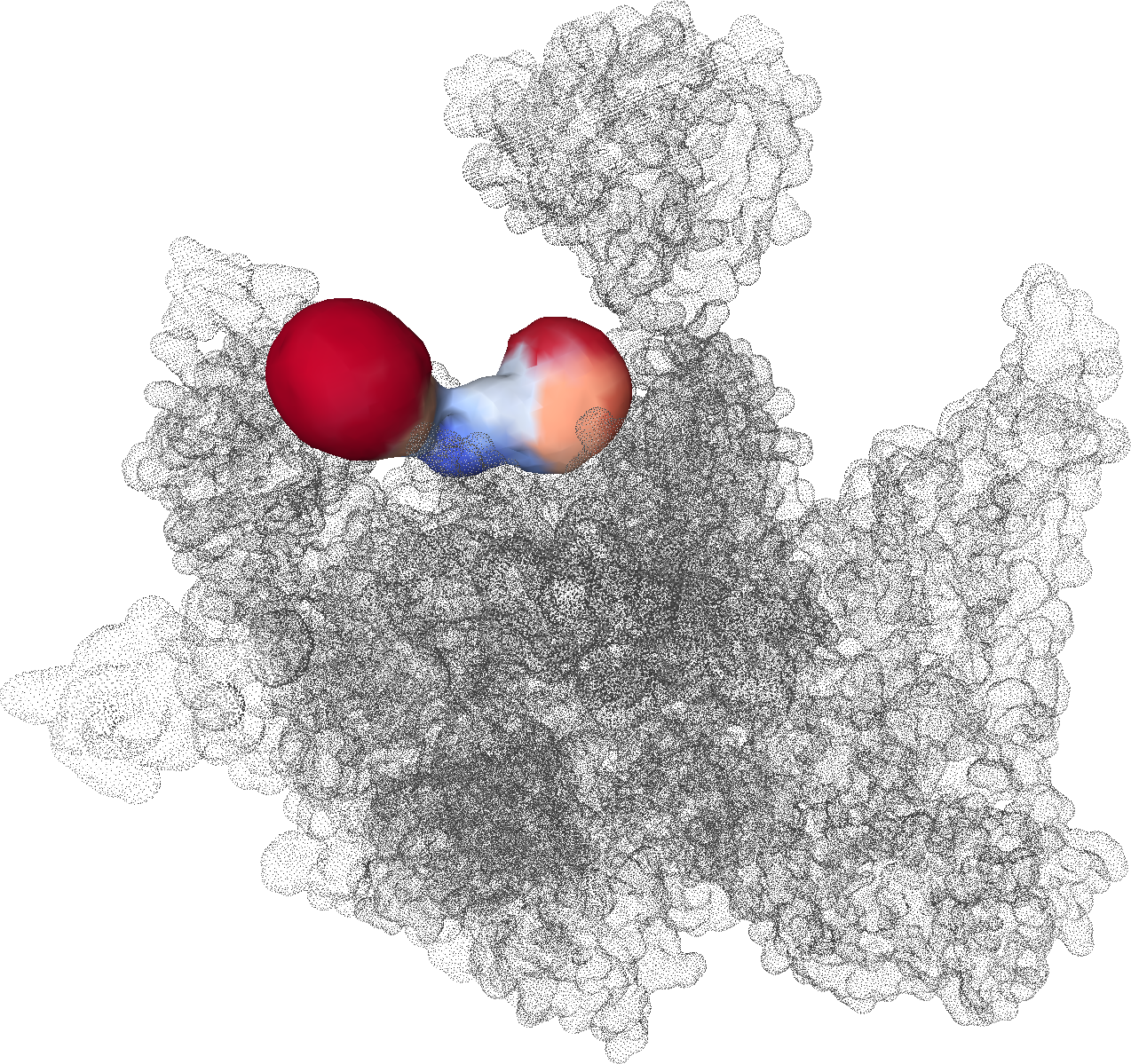}} & \centered{\includegraphics[width=0.19\linewidth]{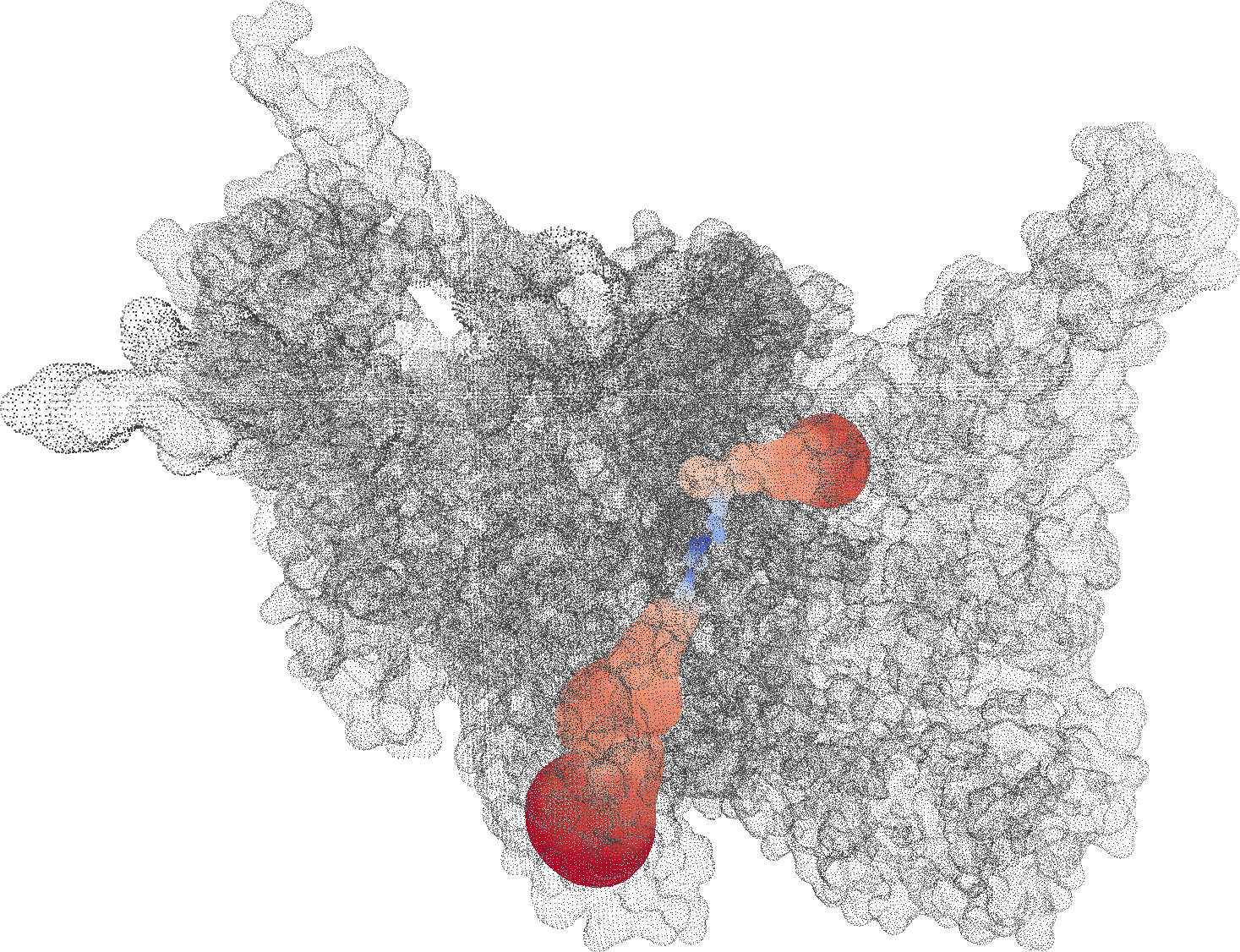}}\\
			\end{tabular}
		}
	\end{center}
	\caption{Centerlines and channels obtained by running  Chanalyzer and HOLE (w.r.t. two different initial positions) on selected dataset instances. Channels are represented as a collection of spheres centred at the points $\mathbf{p}$ of the centerline and of radius $\rho_\mathbf{p}$. Centerlines and sphere points are coloured in accordance with the radius values $\rho_\mathbf{p}$ of the point $\mathbf{p}$ which are in correspondence to. In the pictures, the Solvent Excluded Surface is depicted in grey. The color scheme adopted for representing radius value is the ``coolwarm'' colormap in Matplotlib \cite{Hunter:2007}. It ranges from dark blue for low values to dark red for high values passing from white. In order to improve the visualization as well as the understanding, we decided to normalize the scale to each of the considered examples. Here, the minimum and the maximum values are expressed in $\AA$.
		5EK0: Chanalyzer (min: 0.80, max: 6.09);  1 HOLE (min: 0.58, max: 9.41);   2 HOLE (min: 0.67, max: 9.99).
		7K48: Chanalyzer (min: 1.66, max: 4.14);  1 HOLE (min: 1.43, max: 9.90);   2 HOLE (min: 1.43, max: 10.00).
		7W9K: Chanalyzer (min: 1.84, max: 6.59);  1 HOLE (min: 0.55, max: 9.76);   2 HOLE (min: 3.21, max: 8.62).
		8FHD: Chanalyzer (min: 1.86, max: 5.42);  1 HOLE (min: 0.42, max: 9.42);   2 HOLE (min: 0.45, max: 9.84).}
	\label{fig:comparison}
\end{figure*}

We then proceed with the remaining structures. Figure \ref{fig:comparison} visually depicts -- for some selected structures -- the results obtained by running Chanalyzer and HOLE.
A complete validation is offered in Table \ref{tab:single_analysis}, which collects the measure values of the channels obtained by applying the two tools on the 21 PDB entries of the proposed dataset.

\begin{table*}[h!]\centering
	\caption{Measure values of the channels obtained by running Chanalyzer (Chan.) and HOLE (H.1 and H.2, with respect to two different initial positions) on the proposed dataset. Specifically, we report the number of points of each centerline, their length, their straightness, and the volume of the correspondent channel (represented as the region contained in a collection of spheres).\\}
	\label{tab:single_analysis}
	\resizebox{\columnwidth}{!}{
	\begin{tabular}{|c|ccc|ccc|ccc|ccc|}
		\hline 
		\cellcolor{ForestGreenOriginal!35} &  \multicolumn{3}{c|}{\cellcolor{ForestGreenOriginal!35} $\#$ points} & \multicolumn{3}{c|}{\cellcolor{ForestGreenOriginal!35} Length} & \multicolumn{3}{c|}{ \cellcolor{ForestGreenOriginal!35} Straightness} & \multicolumn{3}{c|}{ \cellcolor{ForestGreenOriginal!35} Volume}\\
		\cellcolor{ForestGreenOriginal!35} PDB code & \cellcolor{ForestGreenOriginal!35} Chan. & \cellcolor{ForestGreenOriginal!35} H.1 & \cellcolor{ForestGreenOriginal!35} H.2 & \cellcolor{ForestGreenOriginal!35} Chan. & \cellcolor{ForestGreenOriginal!35} H.1 & \cellcolor{ForestGreenOriginal!35} H.2 & \cellcolor{ForestGreenOriginal!35} Chan. & \cellcolor{ForestGreenOriginal!35} H.1 & \cellcolor{ForestGreenOriginal!35} H.2 & \cellcolor{ForestGreenOriginal!35} Chan. & \cellcolor{ForestGreenOriginal!35} H.1 & \cellcolor{ForestGreenOriginal!35} H.2 \\
		\cline{1-13}
		
		5EK0 & 237 & 198 & 293 & 47.78 & 94.97 & 99.35 & 0.38 & 0.04 & 0.22 & 1275 & 9299 & 8814 \\ 
		5XSY & 306 & 330 & 179 & 61.69 & 127.55 & 173.91 & 0.16 & 0.07 & 0.05 & 2125 & 10142 & 7203 \\ 
		6AGF & 308 & 335 & 253 & 61.98 & 126.84 & 101.93 & 0.08 & 0.05 & 0.06 & 2898 & 12567 & 9755 \\ 
		6J8E & 316 & 208 & 237 & 63.64 & 130.89 & 124.41 & 0.07 & 0.04 & 0.05 & 2096 & 10194 & 19952 \\
		6J8G & 284 & 306 & 123 & 57.04 & 112.12 & 98.63 & 0.12 & 0.11 & 0.04 & 2405 & 10673 & 8861 \\ 
		6J8H & 296 & 306 & 125 & 59.58 & 114.92 & 89.10 & 0.11 & 0.12 & 0.08 & 2424 & 10304 & 7627 \\
		6J8J & 283 & 127 & 296 & 56.83 & 63.71 & 106.96 & 0.27 & 0.05 & 0.13 & 2438 & 3139 & 10542 \\
		
		6LQA & 294 & 241 & 169 & 59.12 & 119.02 & 101.43 & 0.07 & 0.07 & 0.06 & 2074 & 9178 & 9609 \\ 
		7DTC & 258 & 296 & 316 & 51.96 & 121.66 & 105.04 & 0.23 & 0.05 & 0.12 & 2122 & 13140 & 11639 \\ 
		7DTD & 308 & 226 & 219 & 61.77 & 113.89 & 91.41 & 0.06 & 0.05 & 0.05 & 2656 & 8182 & 10109 \\ 
		7K18 & 266 & 287 & 177 & 53.61 & 95.53 & 100.86 & 0.12 & 0.07 & 0.04 & 2051 & 11616 & 9365 \\
		7K48 & 162 & 252 & 245 & 32.51 & 63.32 & 63.38 & 1.33 & 2.27 & 3.20 & 821 & 8349 & 9088 \\ 
		7W7F & 327 & 124 & 293 & 66.08 & 95.98 & 135.55 & 0.12 & 0.09 & 0.07 & 2102 & 6513 & 12153 \\
		7W9K & 350 & 130 & 63 & 70.49 & 88.50 & 41.76 & 0.10 & 0.04 & 0.06 & 3965 & 6796 & 5978 \\ 
		
		7W77 & 301 & 227 & 200 & 60.60 & 127.63 & 100.28 & 0.11 & 0.05 & 0.04 & 2000 & 7739 & 12902 \\ 
		7WE4 & 267 & 273 & 269 & 53.84 & 101.01 & 102.53 & 0.17 & 0.06 & 0.05 & 1885 & 9800 & 10688 \\ 
		7WEL & 276 & 250 & 289 & 55.53 & 105.78 & 96.32 & 0.24 & 0.05 & 0.07 & 2064 & 11578 & 10277 \\ 
		7WFR & 262 & 210 & 293 & 52.63 & 106.31 & 110.85 & 0.23 & 0.07 & 0.05 & 1956 & 9597 & 11488 \\
		7WFW & 278 & 304 & 277 & 55.87 & 111.68 & 109.41 & 0.24 & 0.05 & 0.07 & 1900 & 9533 & 9886 \\ 
		7XMF & 278 & 290 & 130 & 55.97 & 117.42 & 103.00 & 0.15 & 0.04 & 0.08 & 1744 & 10050 & 8541 \\
		8FHD & 275 & 249 & 191 & 55.38 & 111.55 & 100.90 & 0.13 & 0.06 & 0.06 & 2217 & 6633 & 7099 \\ 
		
		\hline
	\end{tabular}
}
\end{table*}

\begin{figure*}[th!]
	\begin{center}
		\resizebox{0.9\textwidth }{!}{
			\begin{tabular}{c  c} 
				5EK0 & 7K48\\
				\includegraphics[width=0.48\linewidth]{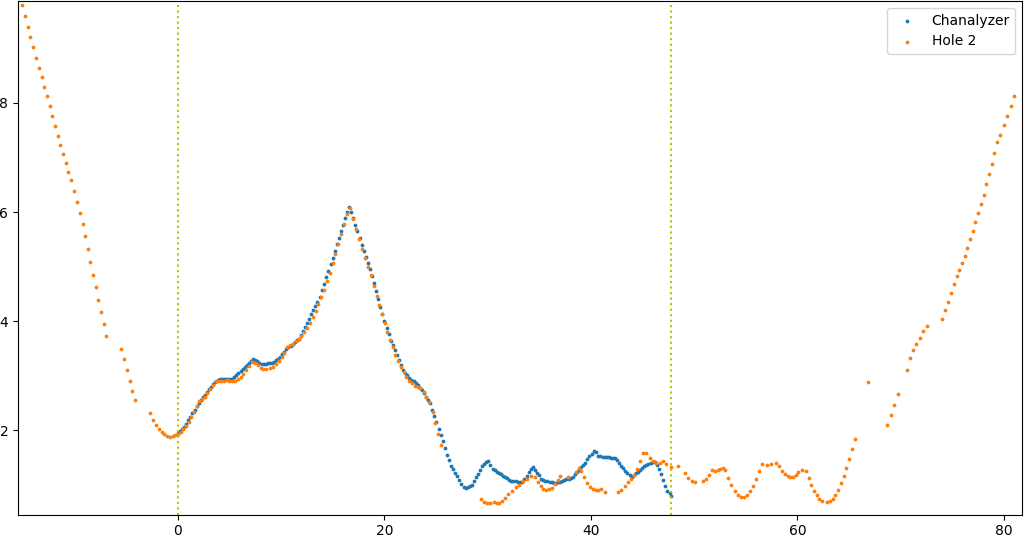} &
				\includegraphics[width=0.48\linewidth]{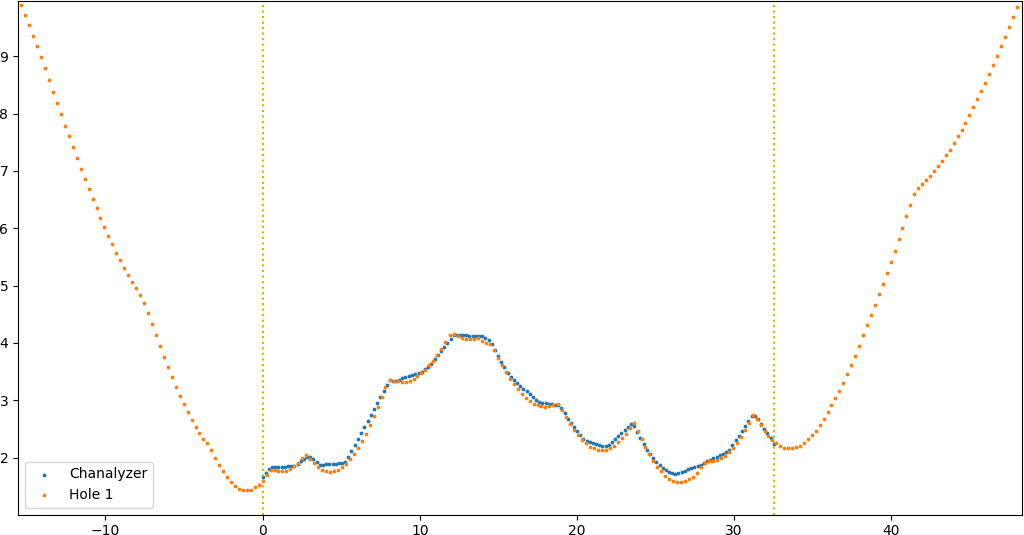}\\
				& \\
				7W9K & 8FHD\\
				\includegraphics[width=0.48\linewidth]{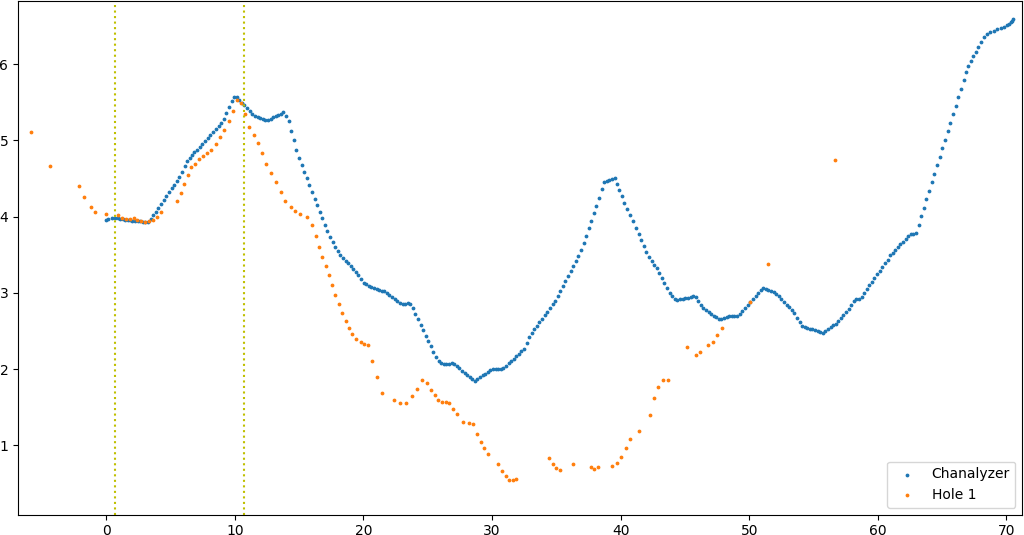} &
				\includegraphics[width=0.48\linewidth]{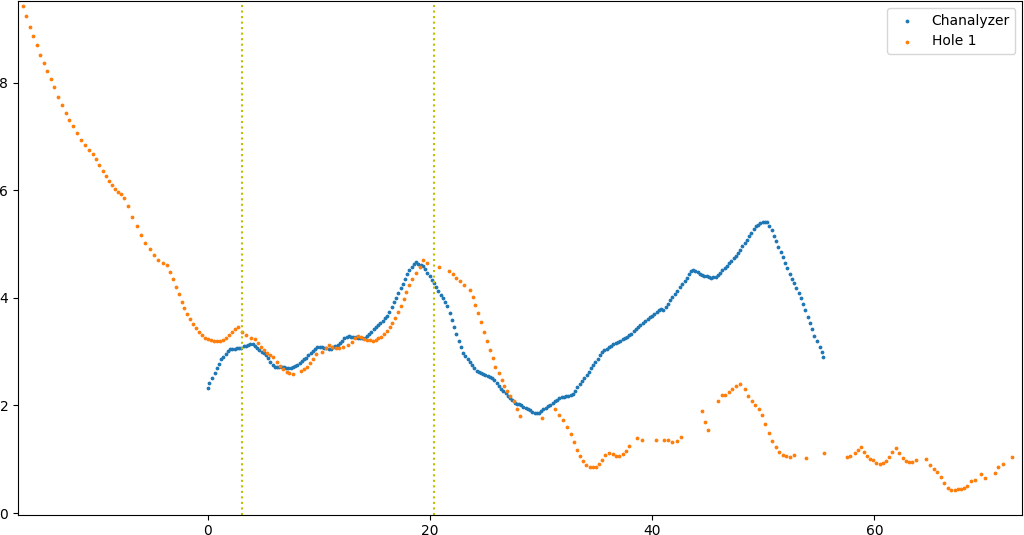}\\
			\end{tabular}
		}
	\end{center}
	\caption{Graphs of the radius functions of the centerlines of the models considered in Figure \ref{fig:comparison}. For each structure, the centerline retrieved by Chanalyzer is depicted in blue while we represent in orange the one among the two produced by HOLE obtaining the higher matching score. Moreover, vertical dashed lines denote the extrema of the interval in which the two centerlines are identified as matched.}
	\label{fig:radius}
\end{figure*}

The values in Table \ref{tab:single_analysis} provide useful information for characterizing the behaviour of the tools Chanalyzer and HOLE. 
In the vast majority of cases, Chanalyzer returns centerlines with lower length and volume values. On the other hand, this tool typically ensures higher values for the straightness parameter. Figure \ref{fig:comparison} gives a visual confirmation of this trend, revealing that HOLE usually produces longer centerlines protruding outward more than the ones retrieved by Chanalyzer which, however, can identify centerlines following a more rectilinear path.
In Figure \ref{fig:comparison} it is also possible to note a marked similarity between the radius values associated with the points of the centerlines received as the output of the two tools; this trend will be confirmed by Table \ref{tab:comparison} and by Figure \ref{fig:radius}.


\begin{table*}[h!]\centering
	\caption{Comparison measure values between the channels obtained by running Chanalyzer (Ch.) and HOLE (H.1 and H.2, with respect to two different initial positions) on the proposed dataset. Specifically, we report in columns $match$ as the percentage of points of a centerline matched with a different centerline. In columns $d_\rho$, we collect the distance between the radius functions of two centerlines on their portion identified as matched.\\}
	\label{tab:comparison}
	\begin{tabular}{|c|cccc|cc|}
		\hline 
		\cellcolor{ForestGreenOriginal!35} &  \multicolumn{4}{c|}{\cellcolor{ForestGreenOriginal!35} $match$} & \multicolumn{2}{c|}{\cellcolor{ForestGreenOriginal!35} $d_\rho$} \\
		\cellcolor{ForestGreenOriginal!35} PDB code & \cellcolor{ForestGreenOriginal!35} (Ch., H.1) & \cellcolor{ForestGreenOriginal!35} (H.1, Ch.) & \cellcolor{ForestGreenOriginal!35} (Ch., H.2) & \cellcolor{ForestGreenOriginal!35} (H.2, Ch.) & \cellcolor{ForestGreenOriginal!35} (Ch., H.1) & \cellcolor{ForestGreenOriginal!35} (Ch., H.2) \\
		\cline{1-7}
		
		5EK0 & 38.40 & 29.80 & 85.23 & 49.15 & 1.88 & 1.94 \\ 
		5XSY & 63.40 & 36.97 & 10.46 & 12.85 & 2.58 & 0.32 \\ 
		6AGF & 49.03 & 34.03 & 23.38 & 20.95 & 1.48 & 1.03 \\ 
		6J8E & 54.11 & 50.48 & 57.59 & 48.10 & 2.85 & 0.69 \\ 
		6J8G & 78.17 & 45.10 & 17.25 & 17.07 & 2.13 & 2.18 \\ 
		6J8H & 51.01 & 30.07 & 9.12 & 14.40 & 2.62 & 1.34 \\ 
		6J8J & 8.48 & 15.75 & 85.87 & 51.35 & 0.54 & 2.15 \\ 
		
		6LQA & 70.75 & 44.81 & 7.14 & 5.33 & 3.90 & 0.53 \\ 
		7DTC & 37.21 & 19.59 & 86.05 & 48.42 & 3.31 & 2.53 \\ 
		7DTD & 30.19 & 24.34 & 21.75 & 18.26 & 0.57 & 1.73 \\ 
		7K18 & 86.09 & 54.36 & 33.83 & 27.12 & 1.61 & 0.68 \\ 
		7K48 & 100.00 & 54.37 & 100.0 & 53.88 & 1.14 & 1.13 \\ 
		7W7F & 2.75 & 3.23 & 44.34 & 25.60 & 0.13 & 3.24 \\ 
		7W9K & 10.57 & 15.38 & 0.00 & 0.00 & 0.38 & 0.00 \\  
		
		7W77 & 28.24 & 20.26 & 29.24 & 23.50 & 2.13 & 1.30 \\ 
		7WE4 & 76.03 & 46.52 & 91.39 & 60.97 & 2.16 & 2.61 \\ 
		7WEL & 80.07 & 53.20 & 87.32 & 57.44 & 3.16 & 2.09 \\ 
		7WFR & 68.32 & 41.90 & 90.08 & 50.85 & 2.83 & 2.80 \\ 
		7WFW & 78.78 & 47.04 & 78.06 & 50.90 & 2.46 & 2.99 \\ 
		7XMF & 78.42 & 47.59 & 11.87 & 13.85 & 4.43 & 0.74 \\ 
		8FHD & 32.00 & 22.89 & 11.64 & 8.38 & 0.59 & 1.11 \\  
		
		\hline
	\end{tabular}
\end{table*}

Table \ref{tab:comparison} collects the values of the comparison measures introduced in Section \ref{sec:measures} for the channels obtained by running Chanalyzer and HOLE on the 21 PDB entries of the proposed dataset. Figure \ref{fig:radius} depicts the graphs of the functions associating each point $\mathbf{p}$ of a centerline with the radius $\rho_\mathbf{p}$. The shown examples concern the models considered in Figure \ref{fig:comparison}. For each model, the considered centerlines are the one retrieved by Chanalyzer and the one among the two produced by HOLE obtaining the higher matching score. Focusing on Table \ref{tab:comparison}, the percentage of matching between the centerlines returned by Chanalyzer and HOLE can assume quite varying values (from a perfect matching for the structure 7K48 to a 0\% matching for 7W9K).
Apart from a few specific cases, the two methods produce output with good matching percentages. 
The values of $match(Ch., \cdot{})$ are almost always greater than the ones of $match(\cdot{}, Ch.)$. This is due to the fact that the centerlines returned by Chanalyzer are typically a portion of the ones retrieved by HOLE.
Moreover, columns $d_\rho$ of Table \ref{tab:comparison} and Figure \ref{fig:radius} reveal that, on the portion they have in common, the centerlines obtained by the two methods are characterized by quite similar radius values.

Table \ref{tab:single_analysis} and Table \ref{tab:comparison} enable us to more rigorously describe the examples depicted in Figure \ref{fig:comparison}.
For structure 7K48, (except for the length) the channel identified by the methods is precisely the same. On the contrary, the channel retrieved by Chanalyzer for model 7W9K does not correspond to any of the two returned by HOLE. For structure 5EK0, there is a good match between the output of Chanalyzer and the channel produced by the second performance of HOLE. Finally, even if on their common part the radius values are pretty similar, the channels identified for model 8FHD have a quite limited matching percentage.

\begin{table*}[th!]\centering
	\small
	\caption{Comparison measure values between the ``exact" structures from the Protein Data Bank repository and the synthetic ones, obtained by running Chanalyzer (exact: Ch.;  synthetic: $\overline{\text{Ch.}}$) and HOLE (exact: H.1 and H.2; synthetic: $\overline{\text{H.1}}$ and $\overline{\text{H.2}}$) with respect to two different initial positions) on the proposed dataset. For the sake of conciseness, the table reports the minimum match and the maximum distance between the pair of centerlines.\\}
	\label{tab:comparison_synth_resume}
	\resizebox{\columnwidth}{!}{
	\begin{tabular}{|c|cc|cc|cc|ccc|}
		\hline 
		\cellcolor{ForestGreenOriginal!35} &  \multicolumn{6}{c|}{\cellcolor{ForestGreenOriginal!35} $\min(match)$} & \multicolumn{3}{c|}{\cellcolor{ForestGreenOriginal!35} $\max(d_\rho)$} \\
		\cellcolor{ForestGreenOriginal!35} PDB code & \cellcolor{ForestGreenOriginal!35} $\left(\,\text{Ch.}, \overline{\text{Ch.}}\,\right)$ & 
		\cellcolor{ForestGreenOriginal!35} $\left(\,\overline{\text{Ch.}}, \text{Ch.}\,\right)$ & \cellcolor{ForestGreenOriginal!35} $\left(\,\text{H.1}, \overline{\text{H.1}}\,\right)$ & 
		\cellcolor{ForestGreenOriginal!35} $\left(\,\overline{\text{H.1}}, \text{H.1}\,\right)$ & \cellcolor{ForestGreenOriginal!35} $\left(\,\text{H.2}, \overline{\text{H.2}}\,\right)$ & 
		\cellcolor{ForestGreenOriginal!35} $\left(\,\overline{\text{H.2}}, \text{H.2}\,\right)$ & \cellcolor{ForestGreenOriginal!35} $\left(\,\text{Ch.}, \overline{\text{Ch.}}\,\right)$ & 
		\cellcolor{ForestGreenOriginal!35} $\left(\,\text{H.1}, \overline{\text{H.1}}\,\right)$ &
		\cellcolor{ForestGreenOriginal!35} 
		$\left(\,\text{H.2}, \overline{\text{H.2}}\,\right)$  \\
		\cline{1-10}
		
		5EK0 & 100.00 & 99.58 & 33.84 & 27.67 & 63.48 & 75.48 & 1.28 & 1.83 & 18.86 \\
		5XSY & 75.16 & 72.27 & 3.94 & 14.81 & 35.20 & 32.68 & 4.68 & 2.18 & 5.50 \\
		6AGF & 83.12 & 81.47 & 20.90 & 34.94 & 0.00 & 0.00 & 5.31 & 3.85 & 0.65 \\
		6J8E & 82.59 & 90.18 & 17.79 & 16.51 & 0.00 & 0.00 & 5.69 & 4.29 & 2.14 \\
		6J8G & 97.54 & 96.55 & 79.74 & 82.24 & 25.20 & 17.72 & 2.07 & 12.70 & 3.99 \\
		6J8H & 94.59 & 69.88 & 10.13 & 16.03 & 0.00 & 0.00 & 4.92 & 7.90 & 2.93 \\
		6J8J & 95.76 & 99.63 & 27.56 & 11.40 & 81.42 & 81.49 & 2.20 & 0.42 & 5.75 \\
		6LQA & 78.91 & 77.55 & 4.98 & 12.87 & 3.55 & 1.84 & 1.83 & 1.52 & 0.35 \\
		7DTC & 93.41 & 90.71 & 8.78 & 40.00 & 97.15 & 95.13 & 3.25 & 7.69 & 15.45 \\
		7DTD & 90.58 & 90.29 & 7.08 & 17.20 & 29.22 & 32.30 & 5.63 & 8.17 & 7.77 \\
		7K18 & 88.72 & 80.55 & 20.91 & 25.93 & 0.00 & 0.00 & 3.93 & 18.16 & 5.70 \\
		7K48 & 100.00 & 99.38 & 95.63 & 94.07 & 96.73 & 96.05 & 0.52 & 14.13 & 14.71 \\
		7W7F & 95.72 & 97.15 & 0.81 & 2.50 & 9.56 & 22.86 & 2.71 & 0.54 & 3.08 \\
		7W9K & 91.71 & 86.27 & 1.54 & 25.00 & 0.00 & 0.00 & 5.07 & 0.66 & 0.00 \\
		7W77 & 77.08 & 70.85 & 28.63 & 29.00 & 0.00 & 0.00 & 3.55 & 0.91 & 1.69 \\
		7WE4 & 90.26 & 89.22 & 26.37 & 44.66 & 83.64 & 76.67 & 4.69 & 10.39 & 12.83 \\
		7WEL & 89.13 & 90.15 & 55.20 & 47.58 & 92.04 & 92.57 & 1.61 & 3.07 & 5.45 \\
		7WFR & 99.62 & 99.61 & 56.67 & 50.97 & 77.47 & 78.52 & 7.30 & 3.06 & 8.83 \\
		7WFW & 72.66 & 72.24 & 15.79 & 18.57 & 80.14 & 81.49 & 2.36 & 6.80 & 3.49 \\
		7XMF & 84.89 & 82.99 & 2.41 & 3.00 & 22.31 & 10.97 & 3.18 & 2.32 & 1.47 \\
		8FHD & 94.91 & 89.49 & 41.77 & 35.60 & 27.75 & 21.36 & 2.63 & 2.35 & 2.04 \\
		\hline
	\end{tabular}
}
\end{table*}

Table \ref{tab:comparison_synth_resume} summarizes the robustness of Chanalyzer and HOLE to local noise. Indeed, for each of the 21 structures retrieved from the Protein Data Bank repository, GEO-Nav is enriched with three additional structures that were created by adding uniform noise locally (see Section \ref{dataset}) and, therefore, it is also possible to assess the geometric robustness of algorithms. In this comparison, the table reports the smallest match values and the largest distances computed between the exact structures and each of the synthetic ones. This test highlights that Chanalyzer is markedly more robust than HOLE, even though some structure point to potential challenges.

\section{Conclusions}\label{sec:conclusions}
In this paper we have presented GEO-Nav, a dataset of 84 voltage-gated sodium channels: $21$ structures originate from the Protein Data Bank Repository, while the remaining $63$ are produced by introducing some uniform random perturbation. To also yield quantitative estimation of the quality of a channel recognition method, we also synthetically created three channels modelled as carbon atoms arranged around an axis to form cylinders of known radius. 

To construct the dataset, we started from Nav channel structures for which the PDB format is known (sometimes there are multiple PDB formats related to the same structure). As a contribution of this dataset, we provide for a good number of Nav channels a molecular surface (namely the SES), offer perturbed surfaces to test the robustness of tools with respect to small perturbations in the atom displacement and provide measurements to evaluate the properties of the recognized channels.

The dataset comes together with several geometric measures that are designed to quantitatively analyse the morphology of the channels, e.g., to study and compare the centerline of each channel and the maximal inscribed ball radius values along it. GEO-Nav has been tested using two methods that are publicly available -- a sphere-based approach and a tessellation-based tool -- showcasing its potential utility for further research and drug design efforts targeting these channels. 

To the best of our knowledge, GEO-Nav is the first dataset that is specifically designed to study the geometric structure of Nav channels.
The availability of a geometry-aware dataset provides valuable resources for advancing the understanding of protein channels in general and, consequently, of their biological function. As a future perspective, we plan to continue to enrich the study with other channels and pores, as well as to consider dynamic features.

\section*{Acknowledgments}
The authors thank: Dr. Walter Rocchia and Dr. Giuseppe Brancato for the useful insights on the geometry of the protein pores and channels; Dr. Giorgio Luciano for the discussions on proteins and the chemist' perspectives on the use of automatic tools for channel identification; Ms. Fabiana Patalano for the fruitful discussions on Nav channels and the help in using the software Vectornator.

This work was carried out within the framework of the projects “RAISE - Robotics and AI for Socio-economic Empowerment” - Spoke number 2 (Smart Devices and Technologies for Personal and Remote Healthcare) and “National Centre for HPC, Big Data and Quantum Computing” HPC - Spoke number 8 (In silico medicine and omics data), and has been supported by European Union - NextGenerationEU.

\bibliographystyle{ieeetr}

\end{document}